\documentclass[
  aps,
  prb,
  twocolumn,
  superscriptaddress,
  floatfix,
  showpacs,
  amsmath,
  amssymb,
  nofootinbib,
  longbibliography
]{revtex4-2}

\usepackage{graphicx}
\usepackage{dcolumn}
\usepackage{bm}
\usepackage{xcolor}
\usepackage{tabularx}
\usepackage{comment}
\usepackage{soul}
\usepackage{csquotes}
\usepackage[normalem]{ulem}
\usepackage{hhline}
\usepackage{braket}
\usepackage{overpic}

\usepackage{hyperref}
\usepackage[capitalize,noabbrev]{cleveref}

\newcommand{\cri}{CrI$_3$}

\newcommand{\tc}{$T_{\rm C}$}
\newcommand{\ts}{$T_{\rm S}$}

\newcommand{\bc}{$B^*$}

\newcommand{\criii}{Cr$^{3+}$}
\newcommand{\sm}{Appendix}

\RequirePackage[normalem]{ulem} 
\RequirePackage{color}\definecolor{RED}{rgb}{1,0,0}\definecolor{BLUE}{rgb}{0,0,1} 
\providecommand{\DIFaddbegin}{} 
\providecommand{\DIFaddend}{} 
\providecommand{\DIFdelbegin}{} 
\providecommand{\DIFdelend}{} 
\providecommand{\DIFaddbeginFL}{} 
\providecommand{\DIFaddendFL}{} 
\providecommand{\DIFdelbeginFL}{} 
\providecommand{\DIFdelendFL}{} 
\newcommand{\DIFscaledelfig}{0.5}
\RequirePackage{settobox} 
\RequirePackage{letltxmacro} 
\newsavebox{\DIFdelgraphicsbox} 
\newlength{\DIFdelgraphicswidth} 
\newlength{\DIFdelgraphicsheight} 
\LetLtxMacro{\DIFOincludegraphics}{\includegraphics} 
\newcommand{\DIFaddincludegraphics}[2][]{{\color{blue}\fbox{\DIFOincludegraphics[#1]{#2}}}} 
\newcommand{\DIFdelincludegraphics}[2][]{
\sbox{\DIFdelgraphicsbox}{\DIFOincludegraphics[#1]{#2}}
\settoboxwidth{\DIFdelgraphicswidth}{\DIFdelgraphicsbox} 
\settoboxtotalheight{\DIFdelgraphicsheight}{\DIFdelgraphicsbox} 
\scalebox{\DIFscaledelfig}{
\parbox[b]{\DIFdelgraphicswidth}{\usebox{\DIFdelgraphicsbox}\\[-\baselineskip] \rule{\DIFdelgraphicswidth}{0em}}\llap{\resizebox{\DIFdelgraphicswidth}{\DIFdelgraphicsheight}{
\setlength{\unitlength}{\DIFdelgraphicswidth}
\begin{picture}(1,1)
\thicklines\linethickness{2pt} 
{\color[rgb]{1,0,0}\put(0,0){\framebox(1,1){}}}
{\color[rgb]{1,0,0}\put(0,0){\line( 1,1){1}}}
{\color[rgb]{1,0,0}\put(0,1){\line(1,-1){1}}}
\end{picture}
}\hspace*{3pt}}} 
} 
\LetLtxMacro{\DIFOaddbegin}{\DIFaddbegin} 
\LetLtxMacro{\DIFOaddend}{\DIFaddend} 
\LetLtxMacro{\DIFOdelbegin}{\DIFdelbegin} 
\LetLtxMacro{\DIFOdelend}{\DIFdelend} 
\DeclareRobustCommand{\DIFaddbegin}{\DIFOaddbegin \let\includegraphics\DIFaddincludegraphics} 
\DeclareRobustCommand{\DIFaddend}{\DIFOaddend \let\includegraphics\DIFOincludegraphics} 
\DeclareRobustCommand{\DIFdelbegin}{\DIFOdelbegin \let\includegraphics\DIFdelincludegraphics} 
\DeclareRobustCommand{\DIFdelend}{\DIFOaddend \let\includegraphics\DIFOincludegraphics} 
\LetLtxMacro{\DIFOaddbeginFL}{\DIFaddbeginFL} 
\LetLtxMacro{\DIFOaddendFL}{\DIFaddendFL} 
\LetLtxMacro{\DIFOdelbeginFL}{\DIFdelbeginFL} 
\LetLtxMacro{\DIFOdelendFL}{\DIFdelendFL} 
\DeclareRobustCommand{\DIFaddbeginFL}{\DIFOaddbeginFL \let\includegraphics\DIFaddincludegraphics} 
\DeclareRobustCommand{\DIFaddendFL}{\DIFOaddendFL \let\includegraphics\DIFOincludegraphics} 
\DeclareRobustCommand{\DIFdelbeginFL}{\DIFOdelbeginFL \let\includegraphics\DIFdelincludegraphics} 
\DeclareRobustCommand{\DIFdelendFL}{\DIFOaddendFL \let\includegraphics\DIFOincludegraphics} 
\RequirePackage{listings} 
\RequirePackage{color} 
\lstdefinelanguage{DIFcode}{ 
  moredelim=[il][\color{red}\sout]{\%DIF\ <\ }, 
  moredelim=[il][\color{blue}\uwave]{\%DIF\ >\ } 
} 
\lstdefinestyle{DIFverbatimstyle}{ 
	language=DIFcode, 
	basicstyle=\ttfamily, 
	columns=fullflexible, 
	keepspaces=true 
} 
\lstnewenvironment{DIFverbatim}{\lstset{style=DIFverbatimstyle}}{} 
\lstnewenvironment{DIFverbatim*}{\lstset{style=DIFverbatimstyle,showspaces=true}}{} 

\begin{document}

\newcommand{\frankfurt}{Institut f\"ur Theoretische Physik, Goethe-Universit\"at Frankfurt,
Max-von-Laue-Strasse 1, 60438 Frankfurt am Main, Germany}

\title{Accessing Few-Layer CrI$_3$ Magnetoelasticity Through Bulk Single Crystals}

\author{Jan~Arneth\footnotemark[*]}
\thanks{Both authors contributed equally to this work.}
\email{j.arneth@kip.uni-heidelberg.de}
\affiliation{Kirchhoff Institute for Physics, Heidelberg University, INF 227, D-69120 Heidelberg, Germany}

\author{Marius Möller\footnotemark[1]}
\thanks{Both authors contributed equally to this work.} \email{mmoeller@itp.uni-frankfurt.de}
\affiliation{\frankfurt}
\author{David A. S. Kaib} %
\affiliation{\frankfurt}
\author{P. Peter Stavropoulos}
\affiliation{\frankfurt}
\author{Aleksandar Razpopov}
\affiliation{\frankfurt}

\author{Martin~Jonak}
\affiliation{Kirchhoff Institute for Physics, Heidelberg University, INF 227, D-69120 Heidelberg, Germany}
\author{Sven~Spachmann}
\affiliation{Kirchhoff Institute for Physics, Heidelberg University, INF 227, D-69120 Heidelberg, Germany}
\author{Mahmoud~Abdel-Hafiez}
\affiliation{Department of Applied Physics and Astronomy, University of Sharjah, P. O. Box 27272 Sharjah, United Arab Emirates}
\affiliation{Physics Department, Faculty of Science, Fayoum University, Fayoum 63514, Egypt}

\author{Sananda Biswas}
\affiliation{\frankfurt}
\author{Kira Riedl}
\affiliation{\frankfurt}
\author{Roser Valent{\'\i}}
\email{valenti@itp.uni-frankfurt.de}
\affiliation{\frankfurt}
\author{Rüdiger~Klingeler} \email{klingeler@kip.uni-heidelberg.de}

\affiliation{Kirchhoff Institute for Physics, Heidelberg University, INF 227, D-69120 Heidelberg, Germany}

\date{\today}

\begin{abstract} 
The persistence of ferromagnetic long-range order in monolayers of the van der Waals semiconductor CrI$_3$ opens new routes for spintronic applications based on two-dimensional quantum magnets. In the fabrication of such devices, the constituent materials inevitably experience anisotropic strain, which modifies their intrinsic electronic properties. At the same time, strain can serve as a powerful tuning parameter, driving the material to desired regimes. While several theoretical studies have investigated the effect of biaxial in-plane strain on CrI$_3$ numerically, experiments are widely limited to the application of hydrostatic pressure. Here, we perform high-resolution magnetostriction experiments on bulk CrI$_3$ samples, and \textit{ab-initio}-based magnetoelastic calculations, to elucidate the role of uniaxial lattice strain on the magnetic properties. Our data show that magnetostriction in CrI$_3$ is unexpectedly sensitive to surface effects, which enables us to investigate the influence of in-plane and out-of-plane strain separately, in both the bulk ferromagnetic (BFM) phase emerging at $T_{\rm C}=61$~K and the surface antiferromagnetic (SAFM) phase below $T^* \simeq 50$~K. In particular, we quantify the uniaxial strain dependence of the surface interlayer coupling $J^{\rm SAFM}_{\perp}$ and the surface spin-flip field \bc, which drastically exceed the strain effects in the BFM phase by a factor of $\sim 30$. The large magnetostrictive response allows us to study the magnetoelastic coupling in few-layer \cri\ through experiments on bulk single crystals, without requiring exfoliation.
\end{abstract}

\maketitle

\onecolumngrid
\section{Introduction}

Among the intensively studied family of quasi-two-dimensional (2D) van-der-Waals (vdW) magnets, the honeycomb trihalide \cri\ is a particularly interesting member, as it is one of the few known materials to exhibit ferromagnetic (FM) long-range order even in the monolayer limit ($T_\mathrm{C}^\mathrm{mono} \simeq 45$)~\cite{huang2017}. In combination with its semiconducting properties, the out-of-plane Ising-like spin ground state renders \cri\ a promising candidate for novel spintronic applications~\cite{marian2023,zhu2020,sierra2021}, with the first nanoscale ``devices'', such as magnetic heterostructures~\cite{hong2025,nnokwe2025,guo2025} and encapsulated carbon nanotubes~\cite{caha2025}, having already been demonstrated.

The observation of ferromagnetic long-range order in a perfectly 2D spin system raises the fundamental question about the origins of the magnetic anisotropy, which is required to circumvent the constraints of the Mermin-Wagner theorem~\cite{merminwagner}. While single-ion anisotropy is not expected to be the dominant source of anisotropy due to the almost fully quenched orbital momentum of the \criii\ ions (spin=3/2), numerical calculations suggest that anisotropic superexchange arising from spin-orbit ($LS$) coupling in the ligand iodine orbitals also plays a major role in stabilizing magnetic order in \cri~\cite{lado2017,kim2019,lee2026,devita2026}. The resulting magnetic anisotropy is strongly uniaxial with an out-of-plane easy $c$ axis, and gives rise to a distinct zone-center gap in the magnon spectrum ($\Delta_\Gamma \simeq 0.33$~meV)~\cite{chen2020,jonak2022}.

In multilayer \cri, the importance of spin-lattice coupling also manifests in the interlayer exchange $J_{\perp}$ strongly depending on the stacking pattern, even changing its sign from antiferromagnetic (AFM) in a monoclinic ($C2/m$) to ferromagnetic in a rhombohedral ($R\overline{3}$) layer stacking~\cite{sivadas2018,jiang2019,soriano2019}. While antiferromagnetic interlayer interaction is found in few-layer systems, bulk single crystals of \cri\ are known to undergo a crystallographic transition ($T_\mathrm{S} \simeq 220$~K) from the high-temperature $C2/m$ to the low-temperature $R\overline{3}$ phase, thus, resulting in the commonly observed ferromagnetic long-range order below $T_\mathrm{C} = 61$~K~\cite{mcguire2015}. However, recent experiments reveal that a mixed surface antiferromagnetic (SAFM) and bulk ferromagnetic (BFM) phase is adopted, where a small fraction of layers proximate to the surface remain in the $C2/m$ structure down to low temperatures and order antiferromagnetically at $T^* \simeq 50$~K~\cite{liu2019,li2020,niu2020}. The magnetic properties of the SAFM phase closely resemble the situation adopted in few-layer \cri~\cite{huang2017,huang2018,niu2020,zhang2021}, which allows for the experimental investigation of the corresponding nanoscale physics in bulk crystals without the need to exfoliate.

When designing and building functional devices based on 2D vdW magnets, the employed materials will unavoidably experience anisotropic strain, for example, due to a lattice mismatch with the used substrate~\cite{bushick2020,yastrubchak2004}. While this may alter the system's favorable intrinsic magnetic behavior, it also offers a route to tune the material's properties in a controlled way~\cite{zhang2021_strain,liu2023_strain}. Therefore, knowledge of the effects of directional lattice distortions on the magnetic properties is a key step to effectively engineer novel spintronic architectures. For \cri, a variety of theoretical studies investigated the influence of biaxial in-plane strain in few-layer systems, suggesting that a compression of the $ab$ plane results in destabilization of the ferromagnetic ground state and eventually even leads to a phase transition into an antiferromagnetic state~\cite{zhang2015,webster2018,liu2018_theo,vishkayi2020}. In contrast, direct experimental investigations are, so far, restricted to the application of hydrostatic pressure. In few-layer \cri, hydrostatic pressure was demonstrated to stabilize the antiferromagnetic ground state, but eventually irreversibly modifies the stacking order at sufficiently high pressure, thereby effectively driving an AFM-to-FM transition~\cite{song2019,li2019,subhan2019}. Similarly, hydrostatic pressure on bulk single crystals initially stabilizes the 2D-ferromagnetism, but very strong compression eventually leads to a full suppression of the ferromagnetic state~\cite{mondal2019,ghosh2022} which is attributed to the change of the Cr-I-Cr bond geometry.

Using thermal expansion measurements on bulk \cri\ single crystals, we previously revealed that the bulk ferromagnetic order can be either stabilized or destabilized depending on whether {\it uniaxial} pressure is applied perpendicular to the chromium layers or along the honeycomb planes, respectively~\cite{arneth2022}. Here, we employ high-precision magnetostriction measurements and \textit{ab-initio}-based magnetoelastic calculations to further elucidate the role of directional lattice strain on the magnetic properties of \cri. Magnetostriction has proven itself as a powerful tool to investigate 2D-van-der-Waals magnets since it allows for the observation of novel hidden phases~\cite{kocsis2022,arneth2025,gass2020}, the calculation of critical pressure dependencies~\cite{gass2020,spachmann2022,spachmann2023} and identification of quantum criticality~\cite{arneth2024,mi2025,mukharjee2026}. Our data reveal pronounced relative length changes associated with the response of antiferromagnetically coupled surface spins, which highlights that magnetostriction in \cri\ is sensitive to both the BFM and SAFM phase. Using thermodynamic analysis, we specifically extract the uniaxial-pressure dependence of the interlayer coupling in the SAFM phase and of the surface spin-flip field \bc .

\section{Methods}
\label{sec:methods}
All experiments were performed on \cri\ single crystals from HQ Graphene\footnote{www.hqgraphene.com}. The crystals were characterized by dc magnetic susceptibility measurements, which confirmed \ts\ $\simeq 212$~K, \tc\ = 61~K, and the antiferromagnetic ordering of the surface layers at $T^*=49.5$~K~\cite{arneth2022,jonak2022}. The dc magnetization and ac magnetic susceptibility were studied by means of a Magnetic Properties Measurement System (MPMS3, Quantum Design). Due to the air sensitivity of the crystals, we did not determine the sample mass prior to the measurements but scaled the isothermal magnetization data to the reported saturation magnetization $M_\mathrm{sat}$ at high fields~\cite{mcguire2015}. High-resolution dilatometry measurements were performed by means of a three-terminal high-resolution capacitance dilatometer in a home-built setup placed inside a Variable Temperature Insert (VTI) of an Oxford magnet system~\cite{kuechler2012,werner2017}. With this dilatometer, the relative out-of-plane and in-plane length changes $dL_\mathrm{i}/L_\mathrm{i}$, i.e., along the crystallographic $c$ axis and in the $ab$ plane, respectively, were studied on a thin single crystal of dimensions $1.2 \times 1.8 \times 0.06~$mm$^{3}$. Measurements of the magnetostriction, i.e., the field-induced length changes $dL_\mathrm{i}(B_\mathrm{i})$, were performed at various fixed temperatures between 1.7~K and 100~K in magnetic fields up to 15~T applied along the direction of measurements.

The strain-dependent spin Hamiltonian for CrI$_3$ was constructed on the basis of first-principles calculations, combining bulk structural relaxations carried out within VASP~\cite{forvasp2} with a Wannier representation obtained using FPLO~\cite{fplo1}. The structural relaxations were carried out for different constrained interlayer distances with fixed value of the conventional $c$ axis, while letting the other lattice constants and atomic positions fully relax. The magnetic exchange couplings were subsequently evaluated for each relaxed structure using the perturbation-theory framework of Ref.~\cite{StavropoulosPRR2021} with $U=5\,\mathrm{eV}$, $J_{\rm H}=1\,\mathrm{eV}$ and $\lambda_p=500\,\mathrm{meV}$ for the I-ligands, while the single-ion anisotropy was obtained by a density functional theory (DFT) total energy-mapping analysis~\cite{riedl2022microscopic}. The \textit{magnetoelastic} couplings $\tilde{\mathcal J}$ were then obtained as the derivatives of each magnetic coupling ${\mathcal J}\in\{J,K,\Gamma,\Gamma',A_c\}$ with respect to $c$-strain $\epsilon_c=\frac{\Delta c}{c_0}$, i.e., $\tilde{\mathcal J}=\frac{\partial \mathcal J}{\partial \epsilon_c}$. To solve the described spin models, we employ a self-consistent cluster mean-field theory method (SCCMFT), in which the \textit{intra}layer interactions ($H_0$, Eq.~\ref{eq:Cri3Hamiltonian}) are handled on a fully-quantum level using exact diagonalization (ED) of eight-site honeycomb clusters, while the \textit{inter}layer interactions ($H_\perp$, Eq.~\ref{eq:H_interlayer}) are incorporated on a mean-field level, $H_\perp\approx J_{\perp} \sum_{\langle ij\rangle_\perp} \mathbf S_\mathrm{i}\cdot \langle\mathbf S_\mathrm{j}\rangle$, where $\langle\mathbf S_\mathrm{j}\rangle$ on neighboring layers are determined in a self-consistency loop. We employ periodic boundary conditions in the in-plane and out-of-plane directions, and use a supercell cluster of two 8-site honeycomb layers, such that the A-type antiferromagnetism can be hosted. For each value and direction of magnetic field $B$, we perform two separate calculations; one for bulk ($J_{\perp}=J^{\rm BFM}_{\perp}=0$) and one for surface layers ($J_{\perp}=J^{\rm SAFM}_{\perp}=0.08\,\mathrm{meV}$), and combine the total result via $\langle \mathcal O\rangle= x \langle \mathcal O\rangle_\mathrm{bulk} + (1-x)\langle \mathcal O\rangle_\mathrm{surface}$, where $x$ denotes the ratio of surface layers. This ratio can be estimated from the initial $M/M_s$ magnetization ratio for small magnetic fields $B\parallel c=B^c_\mathrm{sat}$, c.f.~Fig.~\ref{fig:MS2K_Aniso}(b), such that we estimate $x=0.9$. Further details on the \textit{ab-initio} and SCCMFT calculations are given in the \sm.

\section{Results and Discussion}

\subsection{Experimental Results}
\begin{figure}	
\includegraphics[width=0.5\columnwidth,clip]{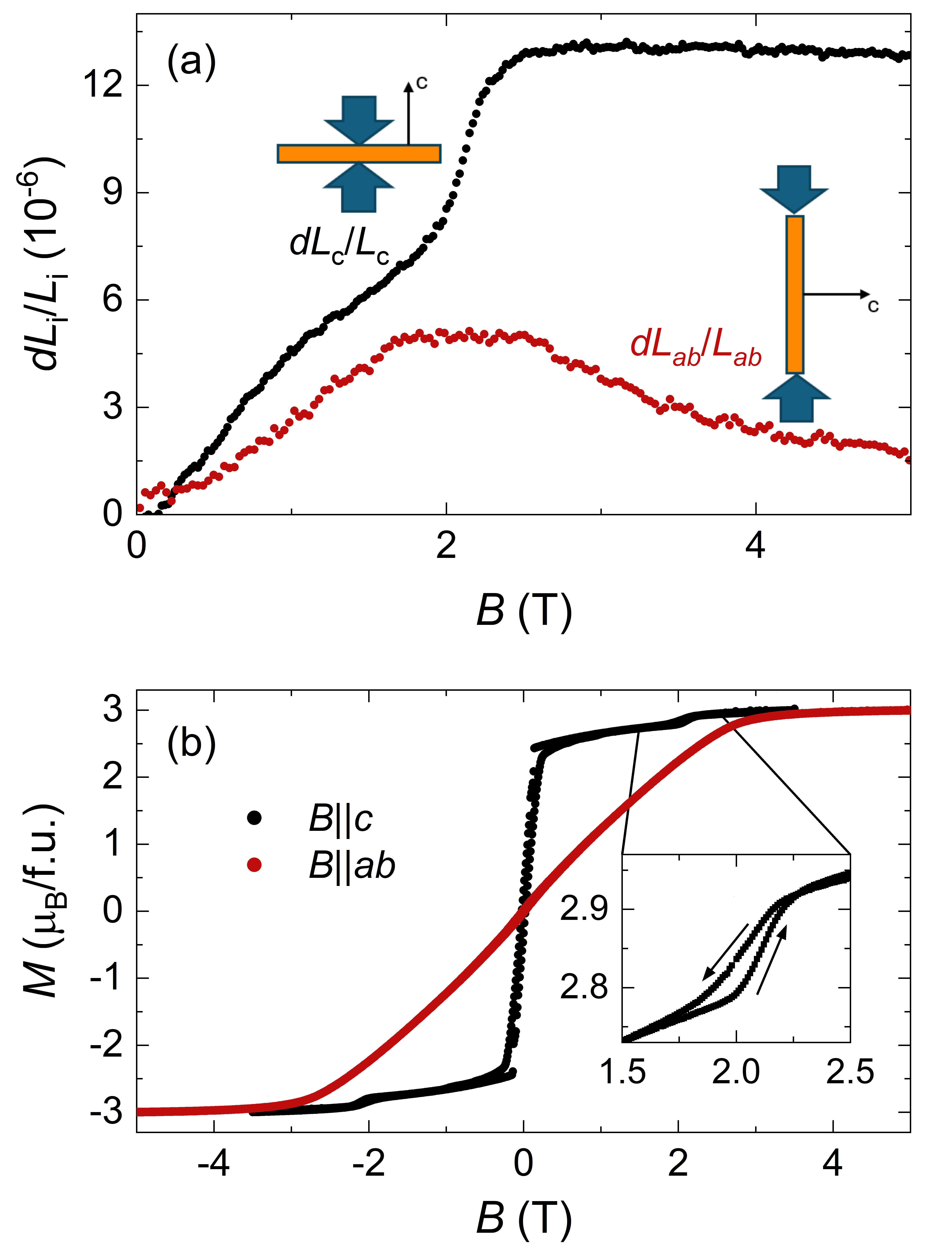}
\caption{(a) Isothermal magnetostriction and (b) magnetization for $B\parallel c$ (black) and $B\parallel ab$ (red) at $T = 2$~K. Sketches in (a) depict the measured length changes. The inset in (b) shows a zoom into the region around \bc.}
\label{fig:MS2K_Aniso}
\end{figure}

The measured isothermal magnetostriction, i.e. the field-induced relative in-plane and out-of-plane length changes, of \cri\ is shown in Fig.~\ref{fig:MS2K_Aniso}(a). Just like the frequently reported isothermal magnetization~\cite{mcguire2015,liu2018,liu2019}, displayed in Fig.~\ref{fig:MS2K_Aniso}(b), the magnetostriction is characterized by pronounced anisotropy. Specifically, the field-induced relative length changes along the $c$ axis are approximately three times larger than those within the $ab$ plane but also exhibit a quite different field-dependence. For $B\parallel c$, $dL_{c}/L_{c}$ initially increases up to 0.23~T, where a slight kink marks the beginning of a quasi-linear regime persisting up to $B^* = 2.1$~T. At $B^*$, the $c$ axis length shows a pronounced jump-like increase, beyond which magnetostriction is virtually constant. In contrast, when the magnetic field is applied along the honeycomb plane, i.e.\ $B\parallel ab$, $dL_{ab}/L_{ab}$ does not show sharp anomalies, but displays a broad plateau-like maximum centered around 2~T. The in-plane magnetostriction coefficient $\lambda_{ab} = \partial L_{ab}/\partial B$  is positive at low fields up to 1.8~T but $L_{ab}$ clearly shrinks for $B > 2.7$~T. These two regimes are separated by the broad plateau-regime mentioned above, where $dL_{ab}/L_{ab}$ only very weakly depends on magnetic field.

\begin{figure*}	[ht]
\includegraphics[width=1\textwidth]{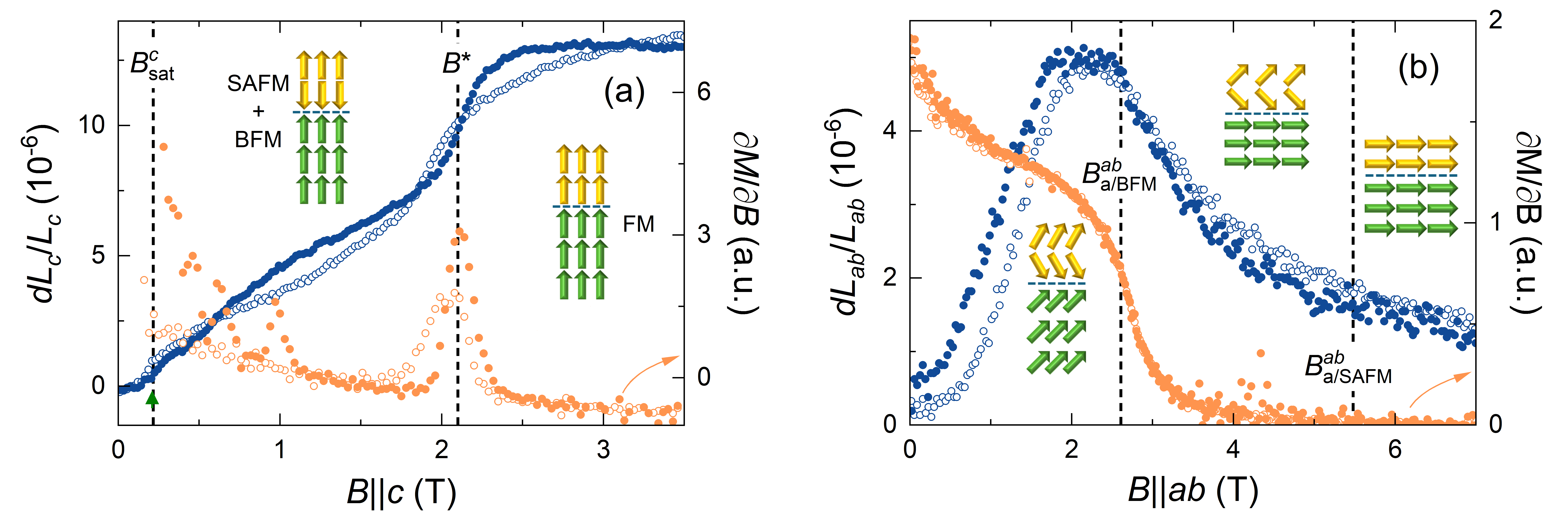}
\caption{Isothermal magnetostriction (blue) and differential magnetic susceptibility (orange) for $B\parallel c$ (a) and $B\parallel ab$ (b) at $T = 2$~K. Vertical dashed lines mark characteristic magnetic fields\footnote{For the determination of $B^{ab}_{\rm a/SAFM}$ see Fig.~\ref{fig:supp_MShighT}. It agrees with the value of $\simeq 5.5$~T found in Ref.~
\cite{mccreary2020}.}, as described in the text. Sketches illustrate the \criii\ spin structure in the corresponding field ranges. Closed (open) circles mark up-(down-)sweeps of the magnetic field.}
\label{fig:ms}
\end{figure*}

The observed magnetostriction in \cri\ can be rationalized by considering the various field-induced spin configurations in the BFM and SAFM phases. As visualized in Fig.~\ref{fig:ms}, the distinct magnetic ground states in each field regime can be distinguished by the features in the relative length changes $dL_\mathrm{i}/L_\mathrm{i}$ and in the differential magnetic susceptibility $\partial M/\partial B$. Already small magnetic fields applied parallel to the magnetic easy $c$ axis give rise to full saturation of the bulk magnetic moments~\cite{mcguire2015,liu2018,liu2019}. Full polarization of the BFM phase is demonstrated in the $M(B\parallel c)$ curves at the saturation field $B_\mathrm{sat}^c = 0.23$~T and is associated with a weak but sharp kink in $dL_{c}/L_{c}$ (green arrow). Despite the full polarization of the BFM phase, the lattice continues to expand even well above $B_\mathrm{sat}^c$, which we attribute to the contribution of \criii\ spins in the antiferromagnetic surface layers that remain unpolarized in this field regime. This assignment is confirmed by the observed pronounced jump in the $c$ axis length at the surface layer spin-flip-transition field $B^* = 2.1$~T~\cite{li2020,niu2020,liu2019}. At $B^*$, the surface layers undergo a first-order phase transition into a spin-flipped phase so that the entire system, i.e., both the BFM and the SAFM phases, assumes a fully polarized ferromagnetic state. Consequently, we do not observe magnetostrictive effects for $B>B^*$.

The assignment of \bc\ to the surface layer spin-flip transition is further corroborated by the absence of the respective anomalies in both the magnetostriction and the magnetization, and by the observation of a negative out-of-plane magnetostriction coefficient $\lambda_c = \partial L_c/\partial B$ in the fully polarised BFM phase ($B>B_\mathrm{sat}^{c}$) at $T \geq 50$~K, i.e., above the surface antiferromagnetic ordering temperature $T^*$ (see Figs.~\ref{fig:supp_MvB_Tdep}(a) and \ref{fig:supp_MShighT}(a)). In agreement to recent studies suggesting that the spin-flip transition at $B^*$ is accompanied by a change in the crystallographic space group of the surface layers from rhombohedral $R\overline{3}$ to monoclinic $C2/m$ ~\cite{li2020}, our measurements provide evidence for a macroscopic magneto-structural response driven by magnetic fields applied parallel to the $c$ axis. Moreover, the observation of a small field-hysteresis of $\Delta B^* \simeq 0.1$~T confirms the discontinuous nature of the spin-flip transition at \bc. Thus, the proposed quantum-critical character of the transition~\cite{li2020} cannot be understood in terms of a conventional continuous quantum critical point. Rather, it would have to correspond to a quantum critical endpoint (QCEP) scenario, in which a line of first-order spin-flip transitions terminates at zero temperature~\cite{garst2005,millis2002,beneke2021}. However, although dilatometry has proven to be a powerful tool for probing QCEPs~\cite{gegenwart2006,weickert2010,arneth2024}, the absence of a discernible anomaly in the thermal expansion coefficient at $T^*$~\cite{arneth2022} does not allow for an assessment of the validity of the proposed quantum nature of the spin-flip transition at $B^*$.

The pronounced uniaxial magnetic anisotropy of \cri\ yields a significant difference of the out-of-plane and in-plane saturation fields in the BFM phase. For $B\perp c$, full polarization of the bulk ferromagnetic moments occurs at the anisotropy field $B^{ab}_\mathrm{a/BFM}= 2.7$~T as compared to $B_\mathrm{sat}^c = 0.23$~T. Concerning the in-plane magnetostriction, the gradual spin alignment into the honeycomb planes is accompanied by an initial expansion of the lattice. The comparison of $dL_{ab}/L_{ab}$ and $\partial M/\partial B$ in Fig.~\ref{fig:ms}(b) indicates that full polarization of the bulk ferromagnetic moments at the anisotropy field $B^{ab}_\mathrm{a/BFM}$ coincides with the right side of the broad magnetostriction plateau. However, the experimental data clearly display a considerable shrinking of the lattice above $B^{ab}_\mathrm{a/BFM}$, where no sizable changes in the isothermal magnetization are visible. Therefore, we associate the finite negative in-plane magnetostriction coefficient with the gradual field-induced polarization of \criii\ spins in the antiferromagnetic surface layers. These spins are not yet fully polarized in this field range and are characterized by a substantially higher anisotropy field $B^{ab}_\mathrm{a/SAFM}$. This assignment is further supported by low-temperature Raman spectroscopy studies on antiferromagnetic few-layer \cri\ reporting anisotropy fields around $\sim 5.5$~T~\cite{mccreary2020}. Thus, in contrast to the bulk magnetization, uniaxial magnetostriction is sensitive to in-plane polarization of the surface layers. Our data, hence, allow us to determine $B^{ab}_\mathrm{a/SAFM}$ from the onset of virtually constant magnetostriction at high magnetic fields. The determination of the anisotropy field is shown in detail in Fig.~\ref{fig:supp_MShighT}(b). At $2$~K, we estimate a surface anisotropy field of 5.8(7)~T, which is in good agreement with the spectroscopically determined value~\cite{mccreary2020}. It should be noted that the corresponding characteristic in $dL_{ab}/L_{ab}$ is not a sharp feature but rather smeared out over a broad field range. We speculate that this may be a result of different surface layers exhibiting slightly different anisotropy fields depending on their distance to the ferromagnetic bulk, which might also rationalize the observation of a small, but finite $B^{ab}_\mathrm{a/SAFM}$ at 50~K $\gtrsim T^*$ (see Fig.~\ref{fig:supp_MShighT}(b)).

The detection of magnetostrictive effects in \cri\ directly insinuates that its magnetic properties can be tuned by external (uniaxial) pressure/stress, which has already been demonstrated experimentally in hydrostatic pressure measurements on bulk and few-layer \cri~\cite{mondal2019,ghosh2023,li2019,song2019}. According to the Maxwell relation 
\begin{equation}
\left.\frac{\partial L_\mathrm{i}}{\partial B_\mathrm{j}}\right|_{p_\mathrm{i}} = -\left.\frac{\partial M_\mathrm{j}}{\partial p_\mathrm{i}}\right|_{B_\mathrm{j}}, \label{max}
\end{equation}
the observed positive out-of-plane magnetostriction coefficient for $B\parallel c$ implies that the macroscopic magnetization decreases upon the application of uniaxial pressure along the $c$ axis. The fact that the BFM and SAFM phase exhibit clearly separated features at different magnetic fields enables us to disentangle and to separately quantify the associated pressure dependencies. Our data reveal a pronounced sensitivity of the lattice to the changes of the spin structure in the SAFM phase, as the jump in $M(B)$ at the SAFM spin-flip field $B^*$ amounts to only $\sim 5~\%$ of the full saturation magnetization, while the corresponding magnetostriction discontinuity accounts for $\sim 40~\%$ of the total measured magnetostriction. This enhanced susceptibility of $dL_{c}/L_{c}(B)$ implies a sizable uniaxial pressure dependence of the critical field \bc , which can be quantified by the Clausius-Clapeyron equation
\begin{equation}
    \frac{\partial B^*}{\partial p_{c}} = V_\mathrm{m} \frac{\Delta (dL_{c}/L_{c})^*}{\Delta M_{c}^*} = 0.50(6) \frac{\rm T}{\rm GPa}.
    \label{Clausius_Clapeyron_BC}
\end{equation}
Here, we used the jumps $\Delta (dL_{c}/L_{c})^* = 4.6(2) \cdot 10^{-6}$ and $\Delta M_{c}^* = 0.15(2)\,\mu_\mathrm{B}/\mathrm{f.u.}$ in the response functions at \bc, as obtained by area-conserving constructions, and the molar volume $V_{\rm m} = 81.3\,\mathrm{cm^3/mol}$. Our analysis reveals that the application of uniaxial pressure along the easy $c$ axis stabilizes the SAFM phase and increases the magnetic field needed to induce the spin-flip into a fully polarized state. 

\begin{figure}	
\includegraphics[width=0.5\columnwidth,clip]{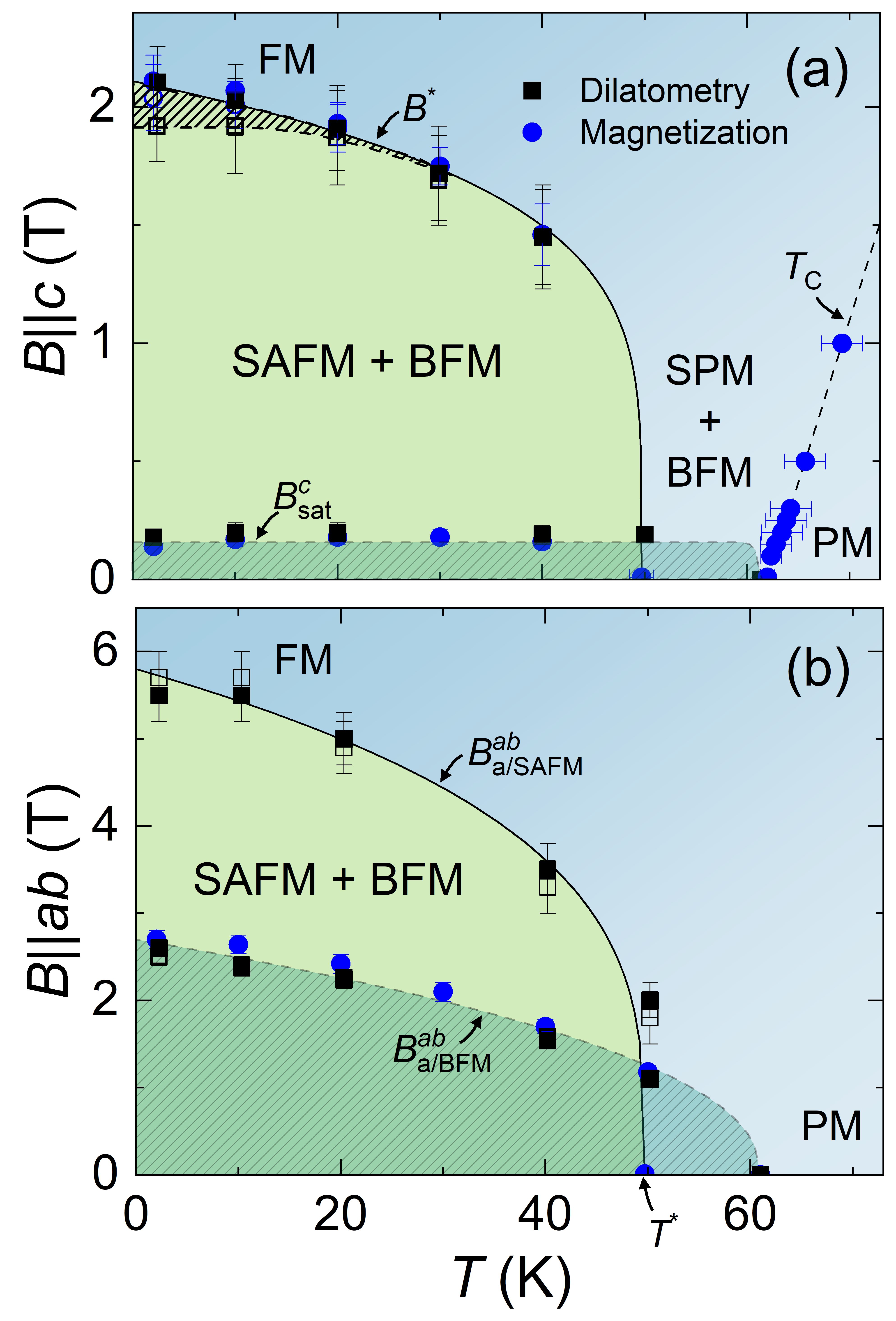}
\caption{Magnetic phase diagrams obtained from dilatometry (black data points) and magnetization (blue data points) measurements for $B\parallel c$ (a) and $B\parallel ab$ (b). PM, FM, SPM+BFM and SAFM+BFM label paramagnetic, ferromagnetic, mixed surface paramagnetic and bulk ferromagnetic, and mixed surface antiferromagnetic and bulk ferromagnetic phases, respectively. In (a), a hysteresis region is denoted by the striped region around \bc. Green shaded areas depict field regimes where the BFM phase is not fully polarized. Data points marking the PM-FM-crossover were obtained by ac susceptibility measurements (see the Appendix for more details). All lines serve as guides to the eye.}
\label{fig:phd}
\end{figure}

Based on Eq.~\ref{max}, the observation of a maximum in $dL_{ab}/L_{ab}(B)$, i.e., the change from a positive to a negative magnetostriction coefficient, implies that there are two regimes with different uniaxial pressure dependencies. Up to $B_\mathrm{a/BFM}^{ab}$, the magnetization will decrease upon compression within the $ab$ plane but $\partial M/\partial p_{ab}$ is positive at higher magnetic fields. Since $B_\mathrm{a/BFM}^{ab}$ marks the full polarization of the ferromagnetic bulk moments into the $ab$ plane, it is straightforward to attribute the measured negative magnetostriction coefficient to the gradual alignment of the antiferromagnetically coupled surface moments. Hence, our data imply that the BFM and SAFM phases exhibit opposite in-plane pressure dependencies and can, therefore, be tuned individually by strain applied within the $ab$ plane. 

Fig~\ref{fig:phd} summarizes the effect of temperature on the characteristic magnetic fields observed in our dilatometric and magnetic measurements in the magnetoelastic phase diagrams for $B \parallel c$ (a) and $B \parallel ab$ (b). The experimental data used to determine these fields at different temperatures are shown in the Appendix. In addition to the results of magnetization and magnetostriction measurements, the field-dependence of the ferromagnetic crossover temperature $T_\mathrm{C}(B\neq 0~\mathrm{T})$, as determined by dynamic magnetic susceptibility studies (see Fig.~\ref{fig:supp_ACChi}), is included in (a). At the lowest temperatures, the phase boundary separating the mixed SAFM + BFM phase from the fully FM state exhibits a weak hysteresis $\Delta B^* \simeq 0.1$~T, thereby confirming its first-order nature~\cite{li2020}. Moreover, \bc\ only weakly depends on temperature in this regime. The slope of the phase boundary at $T = 2$~K and the associated jump in the magnetization allow for an estimation of the associated entropy changes $\Delta S^*$ by employing the Clausius-Clapeyron equation~\cite{Spachmann2021}
\begin{equation}
    \Delta S^* = -\frac{\partial B^*}{\partial T} \cdot \Delta M^* = 8.4(12)\,\mathrm{mJ/molK}.
    \label{Clausius_Clapeyron_DS}
\end{equation}
This result confirms that the spin-flip in the SAFM phase is accompanied by relatively small entropy changes. Analogously, by employing the Clausius-Clapeyron equation of the pressure dependence of the phase boundary, the anomalous length changes at \bc (2~K) from the magnetostriction data in Fig.~\ref{fig:MS2K_Aniso}(a) yield the uniaxial pressure dependence~\cite{Stockert2012} of $T^*$ for $p\parallel c$, i.e.,
\begin{equation}
    \frac{\partial T^*}{\partial p_{c}} = V_\mathrm{m} \frac{\Delta (dL_{c}/L_{c})^*}{\Delta S^*} = 44(7)\,\mathrm{K/GPa}.
    \label{ClausiusClapeyron_T*}
\end{equation}
As the magnetic structure in the surface layers is supposed to be A-type antiferromagnetic~\cite{huang2017}, it is reasonable to assume that the magnetic interlayer coupling $J_{\perp}^{\rm SAFM}$ of the surface layers is the dominant energy scale in driving the phase transition at $T^*$, i.e., $T^* \sim J_{\perp}^{\rm SAFM}$. Hence, we conclude that
\begin{equation}
    \frac{\partial \ln(T^*)}{\partial p_c} = \frac{\partial \ln(J_{\perp}^{\rm SAFM})}{\partial p_c} \simeq 90\%/\mathrm{GPa}.
    \label{eq:pressdepJperp}
\end{equation}
With rising temperature, \bc\ becomes increasingly suppressed  and no corresponding feature can be found in the thermodynamic measurements above $T^*$. At $T = 50$~K, $dL_{c}/L_{c}(B)$ is virtually constant up to $B_\mathrm{sat}^c$ and the out-of-plane magnetostriction coefficient is negative at higher magnetic fields (see Fig.~\ref{fig:supp_MShighT}(a)). As discussed earlier, since the the ferromagnetic bulk moments are fully polarized at $B_\mathrm{sat}^c$, we attribute the shrinking of the $c$ axis to the magnetostrictive response of the \criii\ spins in the paramagnetic surface phase (SPM), that likely remain short-range ordered at temperatures above but close to $T^*$.

Interestingly, the broad maximum in $dL_{ab}/L_{ab}(B\parallel ab)$, marking full saturation of the ferromagnetic bulk, not only shifts to smaller fields but also becomes sharper and evolves into a rather peak-like feature as the temperature is increased (see Fig.~\ref{fig:supp_MShighT}(b)). This directly resembles the observations from isothermal magnetization measurements, where $B^{ab}_\mathrm{a/BFM}$ manifests itself as a point of inflection in $\partial M/\partial B$ at $T = 2$~K but develops a significantly sharper jump-like character with increasing temperature up to \tc\ (see Fig.~\ref{fig:supp_MvB_Tdep}(d)). Similar to the temperature dependence of the bulk anisotropy field, $B^{ab}_\mathrm{a/SAFM}$ also decreases with increasing temperature. Interestingly, a negative magnetostriction coefficient, i.e., the polarization of antiferromagnetic surface layers, is still weakly visible at 50~K. As discussed earlier, we attribute this to a distribution of anisotropy fields in the SAFM phase due to different surface layers exhibiting varying interlayer couplings depending on their distance to the ferromagnetic bulk.

\subsection{Theoretical Modeling}
\begin{figure}[tb]
\includegraphics[width=.4\linewidth]{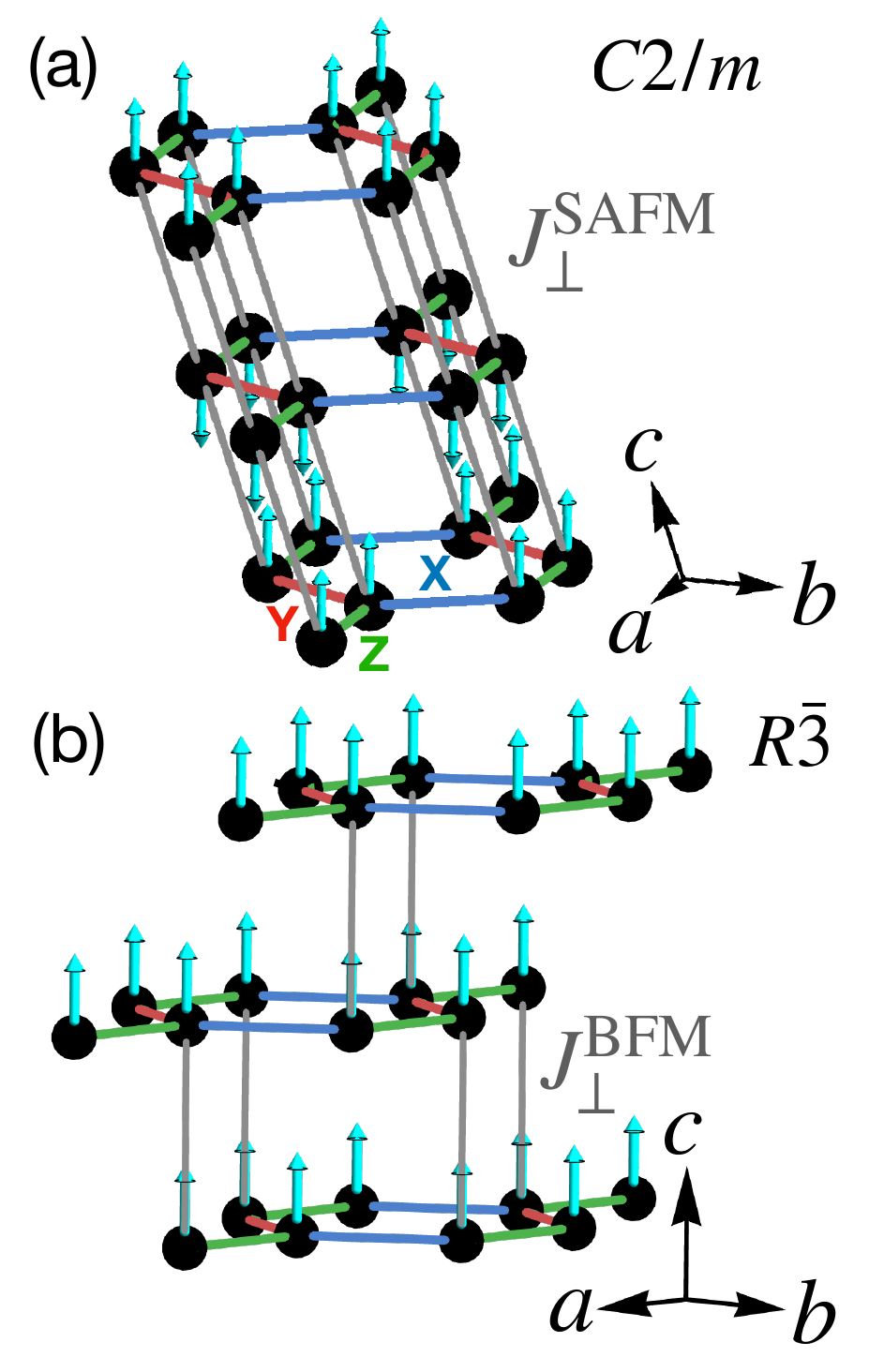}
\caption{Low-temperature Cr site honeycomb network in CrI$_3$ for (a) the layers proximate to the surface (with  $C2/m$ symmetry) and (b) the bulk (with $R\bar{3}$ symmetry). The intralayer nearest neighbors X-, Y- and Z-bonds are show in blue, red, and green respectively. The interlayer bonds are labeled $J_{\perp}^{\rm SAFM}$ and $J_{\perp}^{\rm BFM}$. Each layer consists of 8-sites, representing the cluster used within the SCCMFT calculations.}
\label{fig:CrI3_R3bar}
\end{figure}

To validate the interpretation of our experimental data and to obtain deeper insights into the microscopic origin of magnetostriction, we assume a spin-$\frac{3}{2}$ exchange model of the extended Heisenberg-Kitaev type, including Heisenberg ($J$), Kitaev ($K$) as well as symmetry-allowed symmetric off-diagonal couplings ($\Gamma$, $\Gamma'$), and a single-ion anisotropy (SIA) along the crystallographic $c$ axis ($A_c$) in an external magnetic field ($\mathbf{B}$),
\begin{align}
    H_0 &=  
    \sum_{\gamma, \langle \rm ij\rangle_{\gamma}} \Big[J\, \mathbf{S}_\mathrm{i} \cdot \mathbf{S}_\mathrm{j} +K\, S^{\gamma} _\mathrm{i} S^{\gamma}_\mathrm{j}
    +\Gamma\,(S_\mathrm{i}^\alpha S_\mathrm{j}^\beta + S_\mathrm{i}^\beta S_\mathrm{j}^\alpha) \nonumber\\
    &\qquad \qquad + \Gamma'\,(S_\mathrm{i}^\alpha S_\mathrm{j}^\gamma + S_\mathrm{i}^\gamma S_\mathrm{j}^\alpha + S_\mathrm{i}^\beta S_\mathrm{j}^\gamma + S_\mathrm{i}^\gamma S_\mathrm{j}^\beta) \Big] \label{eq:Cri3Hamiltonian} \\  
    &\quad + A_c  \sum_{i} \left(S_\mathrm{i}^c\right)^2 - g\mu_\mathrm{B}   \sum_{i} \mathbf{B} \cdot \mathbf{S}_\mathrm{i} \nonumber\,, \end{align}
where $\gamma=x,y,z$ corresponds to X-, Y- and Z-bonds as illustrated in Fig.~\ref{fig:CrI3_R3bar} in blue, red, and green, respectively, and $c=\frac{1}{\sqrt3}(x+y+z)$ is the axis perpendicular to the honeycomb plane. 

\begin{table}
\centering
\begin{tabular}{c|cccccc}
\hhline{=======}
$  $ &${A_c}$ & $J$ & $K$ & $\Gamma$ & $\Gamma'$\\ \hline
$\mathcal J$  & $-0.25$ & $-2.01$ & $\phantom{+}0.23$ & $\phantom{+}0.00$ & $\phantom{+}0.00$ \\
$\tilde{\mathcal J}$
& $-1.56$ & $\phantom{+}0.63$ & $\phantom{+}0.36$ & $-0.00$ & $-0.15$ \\
\hhline{=======}
\end{tabular}
\centering
\caption{Zero-strain magnetic $\mathcal J$ and magnetoelastic $\tilde{\mathcal J}=\frac{\partial \mathcal J}{\partial \epsilon_c}$ couplings,in units of meV, with respect to $c$ axis strain $\epsilon_c$, obtained by \textit{ab-initio} calculations in \textit{bulk} CrI$_3$.  Note that the SCCMFT alculations shown later employ an adjusted $A_c=-0.125$\,meV for improved quantitative agreement of critical fields. 
\label{tab:magnetoelasticcouplings}
}
\end{table}

The determined, zero-strain, magnetic ${\mathcal{J}}$ and magnetoelastic $\tilde{\mathcal J}$ couplings for bulk CrI$_3$, obtained as described in the Methods section, are summarized in \cref{tab:magnetoelasticcouplings}. The zero-strain magnetic couplings are in good qualitative agreement with recent studies~\cite{cen2023DeterminingKitaevInteraction}. Interestingly, the magnetoelastic couplings, obey different relative magnitudes and relative signs than the corresponding magnetic couplings, a phenomenon also seen in the Kitaev candidate material $\alpha$-RuCl$_3$~\cite{Kaib2021}. While the zero-strain magnetic $\Gamma'$-exchange is negligible, finite uniaxial strain distorts the octahedral iodine environments, which strongly couples to $\Gamma'$, leading to non-negligible $\tilde {\Gamma'}$ magnetoelastic coupling. Nevertheless, the magnetoelastic couplings of Heisenberg exchange $\tilde J=\partial J/\partial \epsilon_{\rm c}$ and Kitaev exchange $\tilde K=\partial K/\partial \epsilon_{\rm c}$, are also found to be significant, and the SIA magnetoelastic coupling $\tilde A_{\rm c}= \partial A_{\rm c}/\partial \epsilon_{\rm c}$ is found especially strong in comparison. 

While the bulk of CrI$_3$ orders ferromagnetically, the layers proximate to the surface exhibit A-type antiferromagnetism as displayed in Fig.~\ref{fig:CrI3_R3bar}. To capture this three-dimensional magnetic structure, we incorporate an effective interlayer Heisenberg coupling
\begin{equation}
    H_\perp = J_{\perp} \sum_{\langle ij\rangle_\perp} \mathbf S_\mathrm{i}\cdot \mathbf S_\mathrm{j}, \label{eq:H_interlayer}
\end{equation}
to the exchange model $H_0$ where $\langle ij\rangle_\perp$ denotes the nearest-neighbor interlayer bond in the $C2/m$ structure\footnote{\label{fn:Jperp_r3bar_vs_c2m}Since the interlayer coupling $J_{\perp}$ is incorporated classically in the cluster mean-field method, the distinction between the $R\overline{3}$ and $C2/m$ crystal structures effectively amounts to a change of the  parameter $J_{\perp}$ by a constant factor of 2. This is due to the difference in stacking: in the $C2/m$ structure, each Cr site has two nearest-neighbor interlayer bonds, effectively doubling the contribution relative to the $R\overline{3}$ stacking where each Cr site only has one nearest-neighbor interlayer bond, as shown in Fig.~\ref{fig:CrI3_R3bar}.} shown in Fig.~\ref{fig:CrI3_R3bar}. In this description, $J_{\perp}$ represents an effective interlayer coupling constant absorbing other, possibly  further-neighbor or potentially anisotropic, interlayer interaction terms. For the bulk of CrI$_3$, $J_{\perp}=J_{\perp}^{\mathrm{BFM}}$ is expected to be ferromagnetic ($J_{\perp}<0)$ \cite{sivadas2018,jiang2019,soriano2019}. Thus, we have neglected this interlayer interaction in our bulk simulations, as it does not change the bulk ferromagnetic structure. Following the interpretation of the jump at $B^*$ as surface contributions, we incorporate the antiferromagnetic $J_{\perp}=J^\mathrm{SAFM}_{\perp}$ in the $C2/m$ structure and estimate its value by enforcing the model's classical spin-flip field to accord to $B^* \simeq 2.1$~T, which yields
\begin{align}
    J^{\rm SAFM}_{\perp} =  \frac{1}{3} g \mu_\mathrm{B} B^*
    \simeq 0.08\,\mathrm{meV}\,.
\end{align}
By comparing our cluster-mean field results, discussed further below, to experiments, we further estimate its magnetoelastic coupling as $\tilde{J}^{\rm SAFM}_{\perp} = -2.0\, \mathrm{meV}$.

The SCCMFT calculations were performed for both out-of-plane ($B\parallel c$) and in-plane ($B \parallel ab$) directions as described in the Methods section. For the latter case, we find the magnetization and magnetostriction to be qualitatively independent of the in-plane field orientation. Therefore, we present calculations for $B\parallel b$, i.e., magnetic field parallel to a honeycomb bond, as a representative of the general in-plane $B\parallel ab$ response. As all theoretical calculations have been carried out at zero temperature, we compare the numerical results to the experimental data measured at the lowest temperature.

\subsubsection{Magnetization}

\begin{figure}
    \centering
    \includegraphics[width=0.5\linewidth]{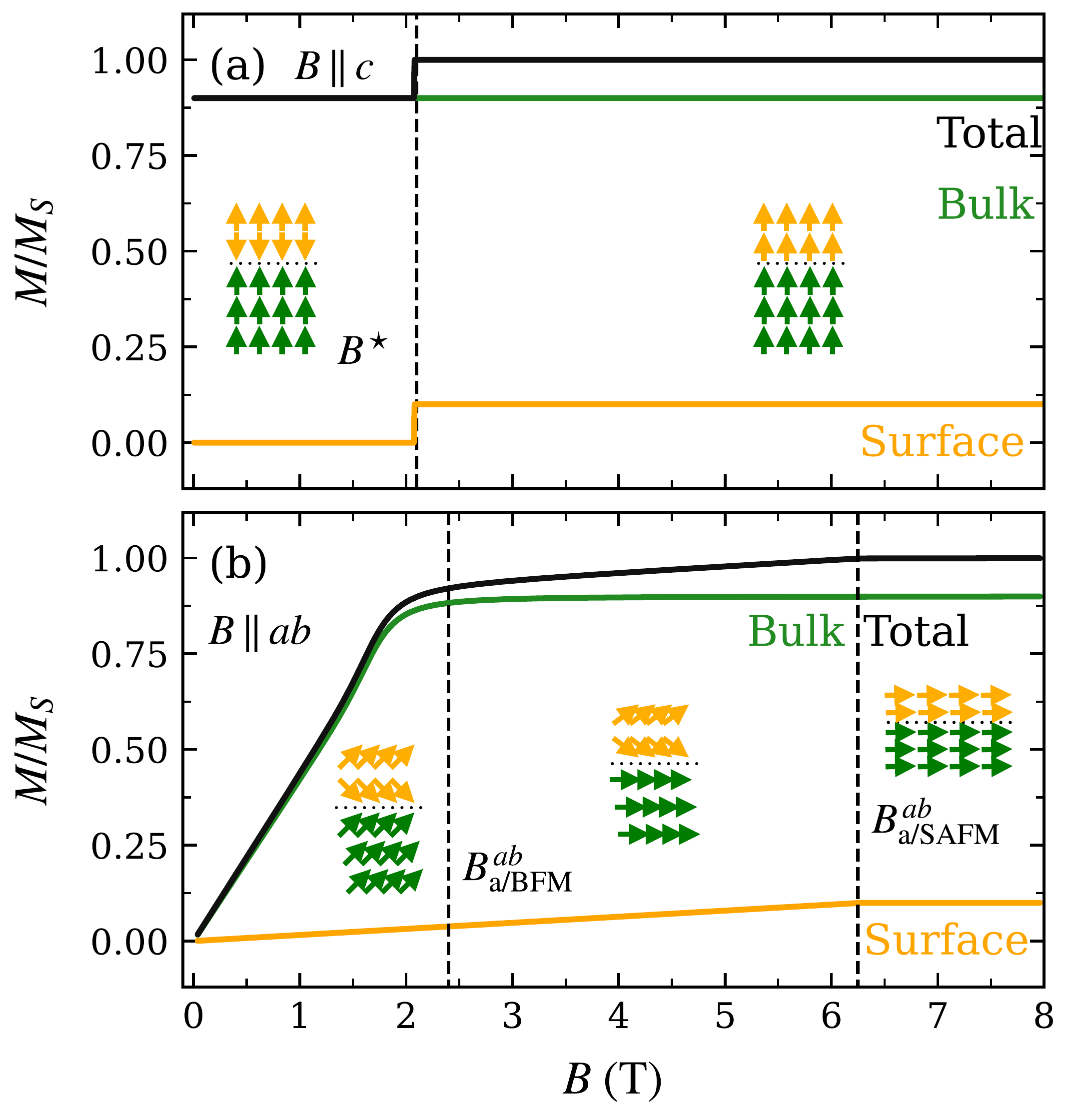}
    \caption{
    Magnetization 
    curves obtained from the self-consistent ED calculation for applied field $B\parallel b$ (a) and $B\parallel c$ (b), split into surface and bulk contributions, in units of saturation magnetization $M_{\rm S}$.}
    \label{fig:MagnetizationCurvesTheory}
\end{figure}

The simulated magnetization curves for magnetic fields applied out-of-plane $B\parallel c$ (a) and in-plane $B\parallel ab$ (b) are displayed in Fig.~\ref{fig:MagnetizationCurvesTheory}. Each plot displays the total magnetization in black, and its constituent contributions from bulk and surface magnetization in green and yellow, respectively. Comparing $M_\mathrm{tot}$ with the lowest temperature $T = 2\,\mathrm{K}$ experimental data displayed in Fig.~\ref{fig:MS2K_Aniso}, we find overall good qualitative agreement. In case of the bulk contribution curves in \cref{fig:MagnetizationCurvesTheory}, the zero-field state is ferromagnetic with an easy $c$ axis, due to the SIA $A_c<0$. Therefore, for $B\parallel c$, the bulk magnetic moments instantly align with applied field for $B\neq0$. 
For $B\parallel ab$ fields, i.e., perpendicular to the easy $c$ axis, the bulk's polarization occurs gradually, as seen for $B\parallel ab$ in Fig.~\ref{fig:MagnetizationCurvesTheory}(b). For the surface contributions to the magnetization, the surface layers exhibit an A-type AFM ground state with no net magnetization and easy $c$ axis at zero field. Increasing the $B\parallel ab$ field, the surface layer $c$ axis moments begin to cant towards the in-plane $ab$ direction, gradually reaching the polarized state at $\sim 6.5\,$T. This is in sharp contrast to the $B \parallel c$ behavior, where the already $c$ axis oriented spins cannot continuously cant towards the magnetic field, and instead undergo a sharp spin-flip transition at $B^\ast\approx 2\,$T. Note that the 2\,K magnetization curve for $B \parallel c$ in experiment is qualitatively smeared out compared to $M_\mathrm{tot}$ from theory. We attribute this discrepancy only partially to finite-temperature effects in the experiment. More importantly, different surface layers are expected to possess distinct internal $B^\ast$ values, likely varying continuously with their distance from the surface. Our model, however, treats the surface contribution using a single fixed $J_{\perp}^{\rm SAFM}$.
\renewcommand{\arraystretch}{1.4}

\subsubsection{Magnetostriction}

Beyond magnetization, we simulated the magnetostriction i.e., the field-induced relative length changes, as measured in the experiment, following the same approximation for the magnetostriction coefficient as in Ref.~\cite{Kaib2021}: 
\begin{align}
        \frac{1}{L_\mathrm{i}}\left. \frac{\partial L_\mathrm{i}}{\partial B_\mathrm{j}}\right|_{p_\mathrm{i}} &= 
        \frac{1}{V} \left. \frac{\partial M_\mathrm{j}}{\partial p_\mathrm{i}}\right|_{B_\mathrm{j}}\nonumber
        \\
        &\approx 
         \frac{\kappa_\mathrm{i} \,L_\mathrm{i} }{V} \sum_{\mathcal J\in\{J,\,K,\, \Gamma',\dots\}}
         \frac{\partial \mathcal J}{\partial L_\mathrm{i}} 
         \left. \frac{\partial M}{\partial \mathcal J} \right|_{B_\mathrm{j}},
\label{eq:magnetostriction_final}
\end{align}
where $V$ is volume, $\kappa_\mathrm{i}=-\frac{1}{L_\mathrm{i}}\frac{\partial L_\mathrm{i}}{\partial p_\mathrm{i}}$ is the linear compressibility against uniaxial pressure $p_\mathrm{i}$, and the sum runs over the magnetic couplings $\mathcal J$ as defined in Eqs.~\eqref{eq:Cri3Hamiltonian}--\eqref{eq:H_interlayer}. The field-dependent relative length change is then obtained by integrating the magnetostriction coefficient, $\frac{dL_\mathrm{i}}{L_\mathrm{i}}=\frac{1}{L_\mathrm{i}}\int_0^B\frac{\partial L_\mathrm{i}}{\partial B_\mathrm{j}}\, \mathrm dB_\mathrm{j}$. Further, we set $\frac{\partial \mathcal J}{\partial L_{c}}=-C \frac{\partial \mathcal J}{\partial L_{ab}}$ where $C$ is a unitless positive constant. This approximation is based on the assumption that compressive (tensile) strain along the out-of-plane ($c$) axis acts, up to a constant scaling factor (C), similarly to tensile (compressive) strain in the ($ab$) plane, which is valid in the linear regime where compression along one direction leads to a proportional expansion in the perpendicular direction.

\begin{figure}
\begin{center}
    \includegraphics[width=0.5\linewidth]{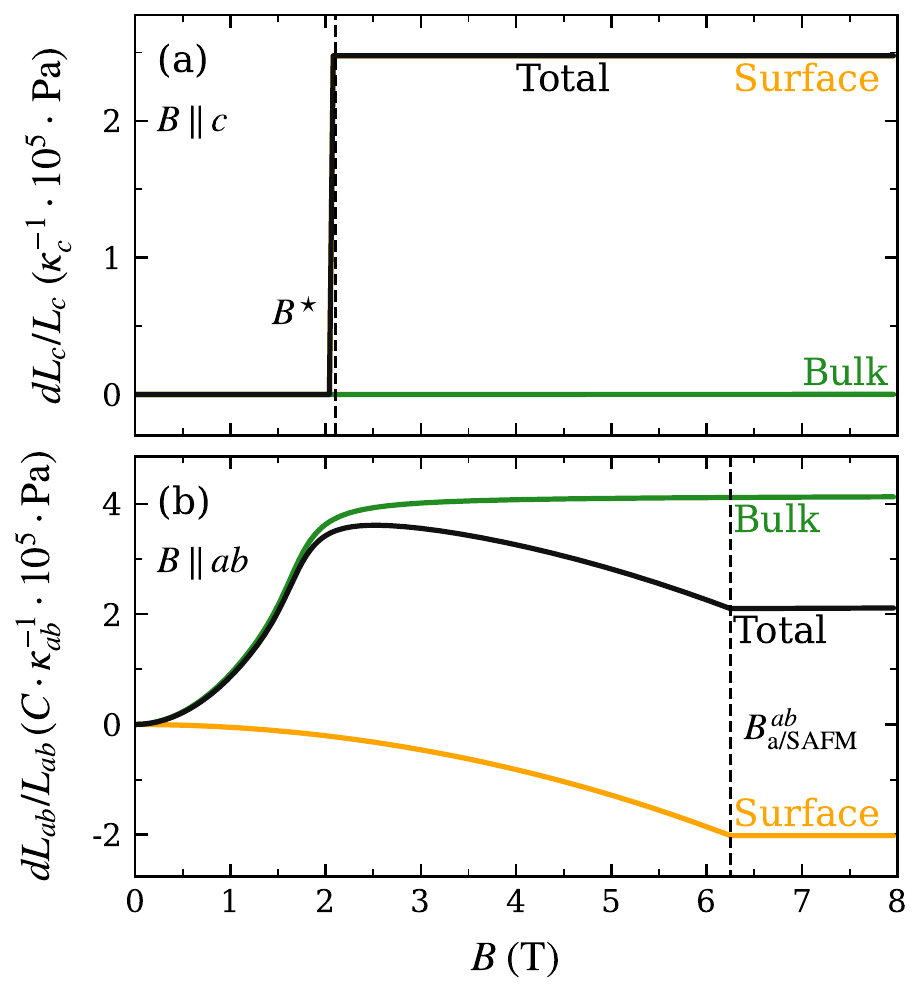}
    \caption{Relative length changes $dL_\mathrm{i}/L_\mathrm{i}$ for applied field (a) along the $c$ axis, and (b) along the $b$ axis (representative $ab$ plane direction), as obtained from SCCMFT, split into surface and bulk contributions.} 
    \label{fig:lambdas_dL_theory}
\end{center}
\end{figure}

Using the discussed formalism, the obtained results for the length change in both field directions are displayed in Fig.~\ref{fig:lambdas_dL_theory}. Focusing on the in-plane magnetic field $B||ab$ (panel (b)), the simulation successfully reproduces both the sign and overall shape of the experimental data. Additionally, it captures the finite $dL_{ab}/L_{ab}$ as observed in the field range of approximately $6.3-8\,\mathrm{T}$. The results indicate that the initial increase in magnetostriction primarily originates from the bulk. Once the bulk magnetization saturates around $B\simeq B^\mathrm{ab}_\mathrm{a/BFM} = 2.7\,\mathrm{T}$, its contribution to the relative length change $dL_{ab}/L_{ab}$ becomes constant. Moving further above $B\simeq B^\mathrm{ab}_\mathrm{a/BFM}$, the ongoing surface response accounts for the subsequent decrease in the total $dL_{ab}/L_{ab}$. When both surface and bulk are saturated, $dL_{ab}/L_{ab}$ remains constant and non-zero, consistent with experimental observations. The response along the in-plane direction is controlled by the interplay of bulk magnetoelastic effects, mainly associated with the SIA $\tilde A_c$ and the anisotropic $\tilde \Gamma'$ term in the bulk and the surface contributions to Eq.~\ref{eq:magnetostriction_final} arising from the interlayer coupling $J_{\perp}^{\rm SAFM}$. Although the magnetoelastic coupling $\tilde{\Gamma'}$ is particularly large, while $\tilde{\Gamma}$ is zero, the qualitative trends can still be reproduced without the anisotropic magnetoelastic contributions, in which case a smaller effective $J_{\perp}^{\rm SAFM}$ is sufficient. The huge contributions of the $\tilde  A_c$ and $\tilde  \Gamma'$ terms are plausible, as those couplings are the most sensitive to the moment re-orientation due to the applied field. A detailed discussion of the individual exchange contributions to the length change can be found in the Appendix \ref{sec:contributionsLengthChange}.

Switching to the length change results for  magnetic field $B \parallel c$, $dL_{c}/L_{c}$ (panel (a)), exhibits a step-like behavior. This step corresponds to a spin-flip transition in a single surface layer at the characteristic field $B^*$. The step originates predominantly from the surface and the interlayer coupling $J_{\perp}^{\rm SAFM}$, with a negligible contribution of other couplings and the bulk in general. The strongly sub-leading bulk response reflects its near-instantaneous polarization for fields applied along the $c$ axis. In the real crystal measured in the experiments, which is composed of multiple layers, each layer may flip at slightly different fields, leading to a smoother, broadened increase observed experimentally.

\subsection{Discussion}
Overall, we find a good qualitative agreement of the theoretically predicted and experimentally measured relative length changes. However, so far, this comparison neglected the absolute magnitude of $dL_{i}/L_{i}$, for which estimates of the uniaxial compressibilities $\kappa_{ab}$, $\kappa_c$, and the scaling factor $C$ are necessary. 
As a reference, we consider the hydrostatic linear compressibilities $k_\mathrm{i}=-\frac{1}{L_\mathrm{i}}\frac{\partial L_\mathrm{i}}{\partial p}$ (versus hydrostatic pressure $p$) of the isostructural compound $\mathrm{CrBr}_3$ from Ref.~\cite{ma16010454}, which below the pressure-induced structural transition were reported as $k_{a} = 0.0067~\mathrm{GPa}^{-1}$ for the in-plane and $k_{\rm c} =  0.0147~\mathrm{GPa}^{-1}$ for the out-of-plane direction. 
As a rough proxy for the corresponding uniaxial compressibilities in $\mathrm{CrI}_3$, we assume values slightly larger than those obtained for $k_\mathrm{i}$ in $\mathrm{CrBr}_3$, based on two considerations.
Firstly, $\mathrm{CrI}_3$ is expected to be more compressible than $\mathrm{CrBr}_3$, consistent with its smaller Young's modulus found when comparing their monolayer variants~\cite{webster2018}. 
Secondly, a uniaxial pressure response, with transverse directions free to relax, is expected to be softer than the corresponding response to hydrostatic pressure; $\kappa_\mathrm{i}>k_\mathrm{i}$.  
We therefore adopt somewhat larger values for CrI$_3$; $\kappa_{ab} \approx 0.01~\mathrm{GPa}^{-1}$, $\kappa_{c} \approx 0.02~\mathrm{GPa}^{-1}$ as order-of-magnitude estimates for the uniaxial compressibilities. 
In addition, we approximate the constant $C$ that connects between the magnetoelastic couplings under $ab$- and those under $c$-strain by comparing the resulting in-plane strain $\epsilon_{ab}$ under enforced $c$ axis strain $\epsilon_{c}$ in our \textit{ab-initio} structural relaxations (see \cref{fig:struct_relx_res}(a)), which yields an estimate of $C\approx - \Delta\epsilon_{ab}/\Delta \epsilon_{c}\approx 0.06$. 

With the described estimates for $\kappa_{\rm ab}, \kappa_{\rm c}$ and $C$, we estimate the absolute magnitude of the maxima for the simulated length changes in both directions. 
For the $ab$ plane calculations of \cref{fig:lambdas_dL_theory}(b), the maximum of the peak is roughly at $3.5\cdot {\mathrm{10^5~Pa}\,C/\kappa_{ab}}$, yielding 
\begin{align}
     (dL_{ab}/L_{ab})_{\rm max} = 3.5 \cdot\mathrm{10^5~Pa}\cdot C^{-1} \cdot \kappa_{ab} \simeq 5.8 \cdot 10^{-5},
\end{align}
which is one order of magnitude higher than in the experimental results. 
For the out-of-plane direction in \cref{fig:lambdas_dL_theory}(a), we find 
\begin{align}
    (dL_{\rm c}/L_{\rm c})_{\rm max} = 2.5\cdot \mathrm{10^5~Pa}\cdot \kappa_{c} = 5 \cdot 10^{-6}, 
\end{align}
which aligns within the experimental order of magnitude. 
Employing the estimates made for $\kappa_{ab}, \kappa_{c}$ and $C$ also allows us to compare the numerically obtained magnetoelastic couplings to the experimentally determined ones. As for the out-of-plane pressure dependence of $J^{\rm SAFM}_\perp$, the values used in our simulations, together with the above estimates, yield $\partial\ln(J^{\rm SAFM}_\perp)/\partial p_{\rm c} \simeq 50\,\%/\mathrm{GPa}$, which is in good agreement with the value estimated from our thermodynamic analysis in Eq.~\ref{eq:pressdepJperp}. Likewise, the \textit{ab-initio} computed isotropic in-plane Heisenberg interaction $J$ and its magnetoelastic coupling $\tilde{J}$, listed in 
\cref{tab:magnetoelasticcouplings}, suggest uniaxial pressure dependencies of $\partial\ln(J)/\partial p_{\rm c} \simeq 0.6\,\%/\mathrm{GPa}$ and $\partial\ln(J)/\partial p_{\rm ab} \simeq -5.2\,\%/\mathrm{GPa}$. Despite differing in magnitude, the signs of these values are in qualitative agreement to those obtained from Grüneisen analysis of the thermal expansion data,  i.e., $\partial\ln(J)/\partial p_{\rm c} = 2.8(5)\,\%/\mathrm{GPa}$ and $\partial\ln(J)/\partial p_{\rm ab} = -0.7(2)\,\%/\mathrm{GPa}$, as reported by us earlier~\cite{arneth2022}.
Considering that the strength of magnetostrictive effects can vary from material to material by multiple orders of magnitudes as well as the qualitative estimates we have made for this conversion, and that the significant surface layers actually likely have different elastic properties, we find the level of agreement between our \textit{ab-initio} based calculations and the experiments on bulk single crystals reaffirming.

The magnetic response of \cri\ to lattice distortions has been directly experimentally investigated in several hydrostatic pressure studies on both bulk~\cite{mondal2019,ghosh2022} and few-layer~\cite{li2019,song2019} samples. In general, hydrostatic pressure dependencies comprise the sum of the distinct orthogonal uniaxial pressure dependencies, which prevents the direct comparison of our results with the reported values. However, due to the strong structural anisotropy of \cri, the effects of hydrostatic pressure are expected to be dominated by the response of the $c$ axis, so that the uniaxial-pressure response along the $c$ axis at least roughly approximates the hydrostatic-pressure effects~\cite{li2019}. This has been experimentally confirmed by recent thermal expansion studies~\cite{arneth2022}. In this respect, the observed overall positive out-of-plane magnetostriction coefficient and, hence, the concluded decrease of the magnetization upon compression of the $c$ axis, are in qualitative agreement with hydrostatic pressure studies on bulk \cri~\cite{mondal2019,ghosh2022}. Moreover, the calculated positive out-of-plane pressure dependence of the SAFM ordering temperature $T^*$ (Eq.~\ref{ClausiusClapeyron_T*}) not only confirms the hydrostatic pressure results\footnote{Note, that our experimental uniaxial pressure dependencies of $T^*$ have been determined at $B \simeq 2.1$~T while Mondal~{\it et al.} performed their measurements at $B=0.05$~T.} of Ref.~\cite{mondal2019} but implies that antiferromagnetic long-range order in the surface layers is strongly stabilized by uniaxial pressure along $c$. The stabilization of the SAFM phase due to out-of-plane pressure also becomes evident through the positive value of $\partial B^*/\partial p_{c}$ (Eq.~\ref{Clausius_Clapeyron_BC}). A weak increase of the spin-flip field under hydrostatic pressure has also been demonstrated in few-layer \cri\ experiments, which, moreover, reveal an irreversible pressure-induced AFM to FM transition, connected to a switching of the layer stacking order, above $p \simeq 1.4$~GPa~\cite{li2019,song2019}.

While the experimental demonstration of in-plane strain effects in \cri\ is restricted to our previous thermal expansion measurements of the ferromagnetic bulk phase from which $\partial T_{\rm C}/\partial p_{\rm ab}$ was determined~\cite{arneth2022}, several numerical studies investigated the response of few-layer systems to $ab$ lattice distortions theoretically. In the monolayer, compressive biaxial strain is suggested to destabilize the ferromagnetic ground state through a decrease of the nearest-neighbor Heisenberg exchange $J_1$ and to eventually result in a phase transition into an antiferromagnetic state~\cite{zhang2015,webster2018,liu2018_theo,vishkayi2020}. Likewise, shrinking of the $ab$ lattice plane in bilayer \cri\ is reported to modify the layer stacking order and, thereby, weaken the antiferromagnetic ground state~\cite{leon2020,xu2021,ouyang2025}. The reported numerical predictions are consistent with our experimental data, as both of the observed pressure dependencies, i.e., $\partial M/\partial p_{ab} < 0$ in the BFM and $\partial M/\partial p_{ab} > 0$ in the SAFM phase, signal the destabilization of the respective magnetic order under in-plane pressure. In particular, the strong stacking-dependence of $J_{\perp}$ in \cri~\cite{sivadas2018,song2019,zhang2021} renders comparison of bulk and surface magnetism a complicated task, as the two phases are expected to exhibit different crystal structures~\cite{li2020}. However, it is likely that a pressure-induced decrease of the interlayer spacing leads to an increase of the overlap between the wave functions which are involved in the dominant higher-order superexchange processes~\cite{sivadas2018}.

When comparing the uniaxial pressure dependencies of the SAFM and BFM ordering temperatures, i.e. $\partial \ln(T^*)/\partial p_{\rm c} \simeq 90\,\%/\mathrm{GPa}$ and $\partial \ln(T_\mathrm{C})/\partial p_{\rm c} \simeq 3\,\%/\mathrm{GPa}$, as determined here and in an earlier study~\cite{arneth2022}, respectively, our data reveal that the \cri\ surface layers are by a factor of $\sim 30$ more sensitive to uniaxial pressure than the bulk layers. Thus, we observe pronounced features related to surface spins even in the magnetostriction of bulk \cri\ samples. Since the microscopic properties in the SAFM phase of bulk crystals resemble those in few-layer samples~\cite{huang2017,liu2018,niu2020,li2020}, our measurements provide a route to experimentally study the uniaxial strain dependencies of thin-film \cri\ on macroscopic samples without the need of exfoliation.

\section{Summary}

In conclusion, we find that magnetostriction in bulk \cri\ is astonishingly sensitive to surface magnetism and use this tool to investigate and quantify the effects of uniaxial pressure in both the ferromagnetic bulk phase and the antiferromagnetic surface phase. For magnetic field applied perpendicular to the honeycomb planes $B \parallel c$, the spin-flip transition into a fully ferromagnetic state at the surface is accompanied by a pronounced jump in the $c$ axis length. Thermodynamic analysis shows that the critical field can be effectively tuned at a rate of $\partial B^*/\partial p_\mathrm{c} = 0.50(6)\,\mathrm{T/GPa}$ by external uniaxial pressure. Magnetic fields applied along the honeycomb plane $B||ab$ yield a positive  magnetostriction coefficient in the BFM phase and a negative one in the SAFM phase. The distinct features in $dL_{ab}/L_{ab}$ allow us to follow the temperature dependence of both the bulk and the surface in-plane saturation field. We construct the magnetoelastic phase diagram from which we deduce that uniaxial pressure along the $c$ axis  stabilizes interlayer coupling in the antiferromagnetic surface phase at a rate $\partial \ln(J_{\perp}^{\rm SAFM})/\partial p_\mathrm{c} \simeq 90\,\%/\mathrm{GPa}$. Our data, hence, elucidate the importance of spin-lattice coupling in quasi-2D van-der-Waals materials, quantify the effects of surface magnetism in bulk samples, and specifically provide valuable information on the uniaxial strain effects in antiferromagnetic few-layer \cri .

\begin{acknowledgements}
J.A. and R.K. acknowledge support by the Heidelberg-Karlsruhe Strategic Partnership and the Heidelberg University's Cluster of Excellence STRUCTURES EXC2181/1-390900948. J.A. acknowledges support by the IMPRS-QD Heidelberg. {M.M., D.A.S.K., P.P.S., A.R., S.B., K.R., and  R.V. gratefully acknowledge support by the {Deutsche Forschungsgemeinschaft (DFG, German Research Foundation) for funding through TRR 288-422213477 (projects A05, B05).}}
\end{acknowledgements}

\appendix

\renewcommand{\thefigure}{A\arabic{figure}}
\setcounter{figure}{0}  

\section{Experimental Results}\label{sec:a1}
\subsection{Magnetostriction}
\begin{figure}[h]
    \centering
    \includegraphics[width = 0.5\columnwidth]{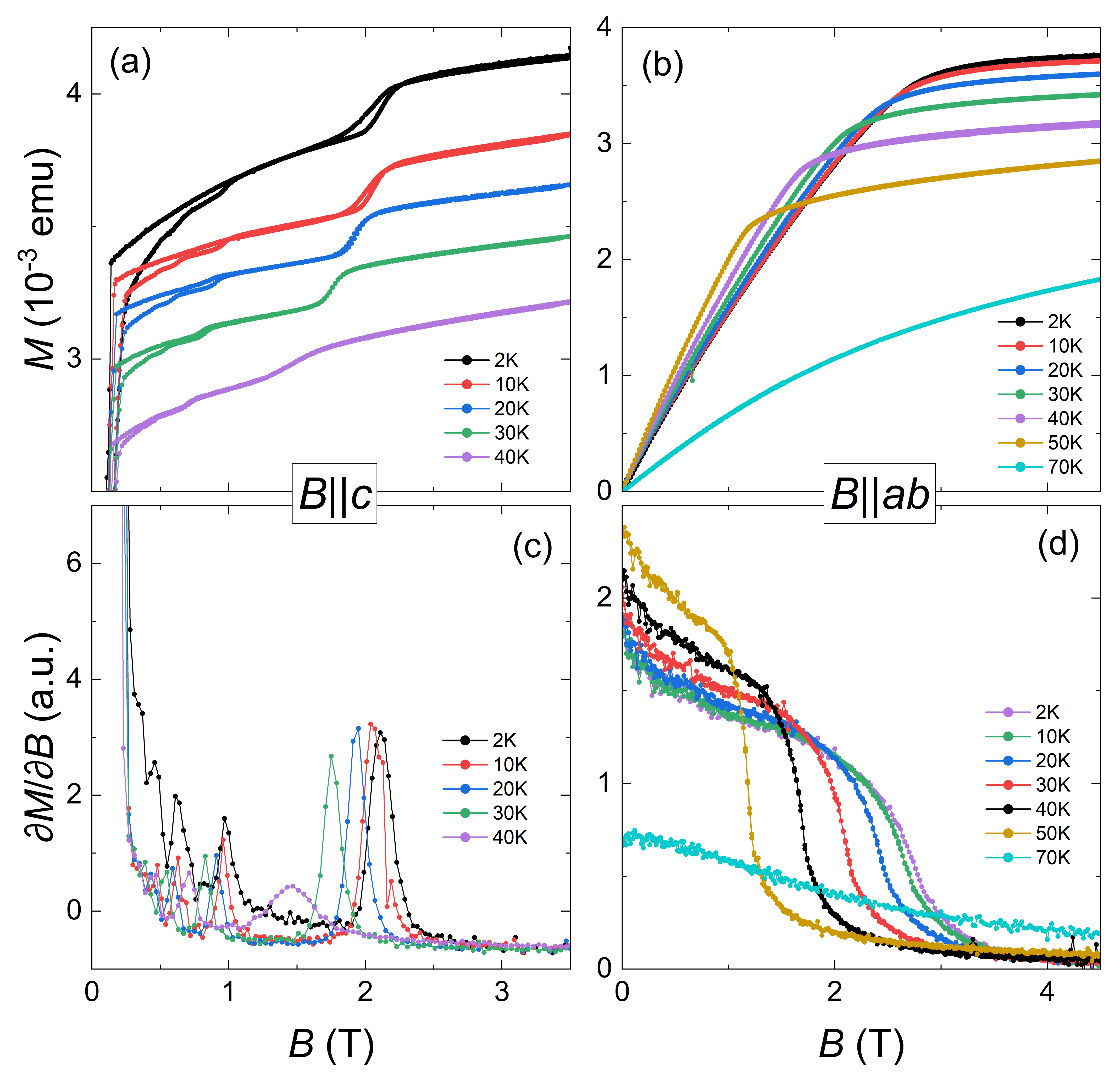}
    \caption{Isothermal magnetisation (upper row) and differential susceptibility (lower row) for $B||c$ (left column) and $B||ab$ (right column) at selected temperatures. In (c) only up-sweeps of the magnetic field are shown to increase visibility.}
    \label{fig:supp_MvB_Tdep}
\end{figure}

\begin{figure}
    \centering
    \includegraphics[width = 0.5\columnwidth]{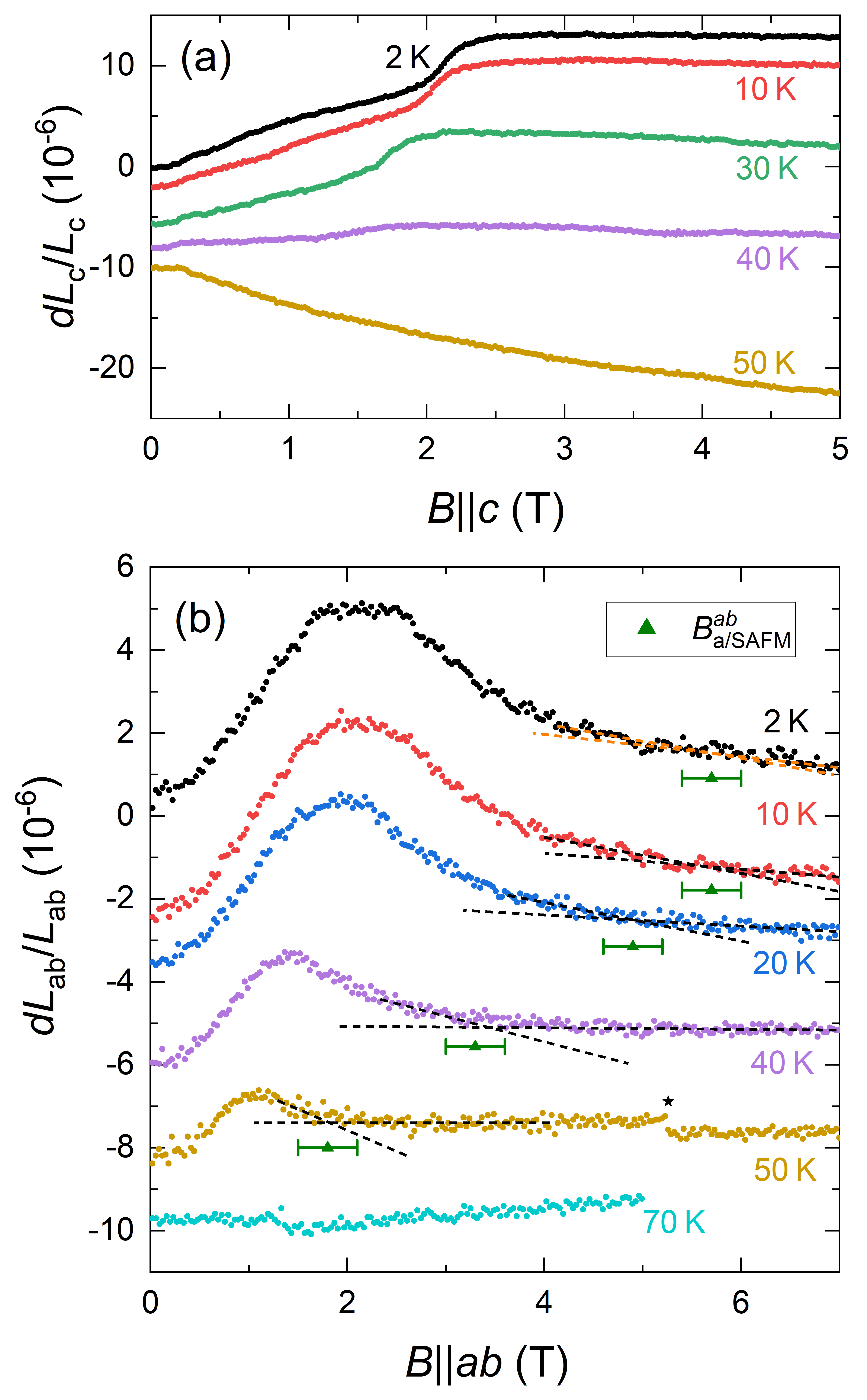}
    \caption{Isothermal magnetostriction for $B\parallel c$ (a) and $B\parallel ab$ (b) at various temperatures. For visibility, only up-sweeps of the magnetic field are shown and the data are shifted by $2\cdot 10^{-6}$ along the ordinate. Dashed lines indicate the determination of $B^{\rm ab}_{\rm a/SAFM}$, as marked by green triangles. The asterisk marks an experimental artifact.}
    \label{fig:supp_MShighT}
\end{figure}

The isothermal magnetization and corresponding (differential) magnetic susceptibility of \cri\ at various temperatures is shown in Fig.~\ref{fig:supp_MvB_Tdep} for $B\parallel c$ (a,c) and $B\parallel ab$ (b,d). As for $B\parallel c$, the spin-flip transition in the SAFM phase is suppressed and occurs at smaller fields $B^*$ as the temperature increases. Furthermore, the hysteresis at \bc\ is diminished and no considerable difference between the magnetic field up- and down-sweeps is observed at $T>40$~K. Notably, the magnetic field up-sweeps exhibits several weak discontinuities before the BFM phase reaches full saturation, leading to a pronounced hysteresis at $B \leq 1.1$~T, too. Similar behavior has been recently observed in the magnetoresistance measurements of \cri-based tunneling junctions, where its occurrence was attributed to a successive flipping of pinned individual layers or domains~\cite{wang2018}.
Fig.~\ref{fig:supp_MShighT} shows the effect of temperature on the out-of-plane (a) and the in-plane (b) magnetostriction. In general, the magnetostrictive response is found to decrease upon heating, which can be attributed to the interplay of thermal fluctuations and the magnetocrystalline anisotropy. As expected, the temperature dependencies of the SAFM spin-flip field \bc, as marked by pronounced discontinuous lattice changes in $dL_c/L_c$, and of the bulk anisotropy field $B^{ab}_\mathrm{a/BFM}$, marked by the onset of negative magnetostriction in $dL_{\rm ab}/L_{\rm ab}$, closely resemble the behavior observed in the magnetization measurements. The data imply negative out-of-plane magnetostriction coefficients at $T=50$~K above $B^c_\mathrm{sat}$, which we attribute to the response of the paramagnetic surface layers above $T^*$. The inset in Fig.~\ref{fig:supp_MShighT}(b) depicts the determination of the SAFM anisotropy field $B^{ab}_\mathrm{a/SAFM}$ from the onset of virtually zero (linear) magnetostriction.

\subsection{Magnetic susceptibility}

\begin{figure}
    \centering
    \includegraphics[width = 0.5\columnwidth]{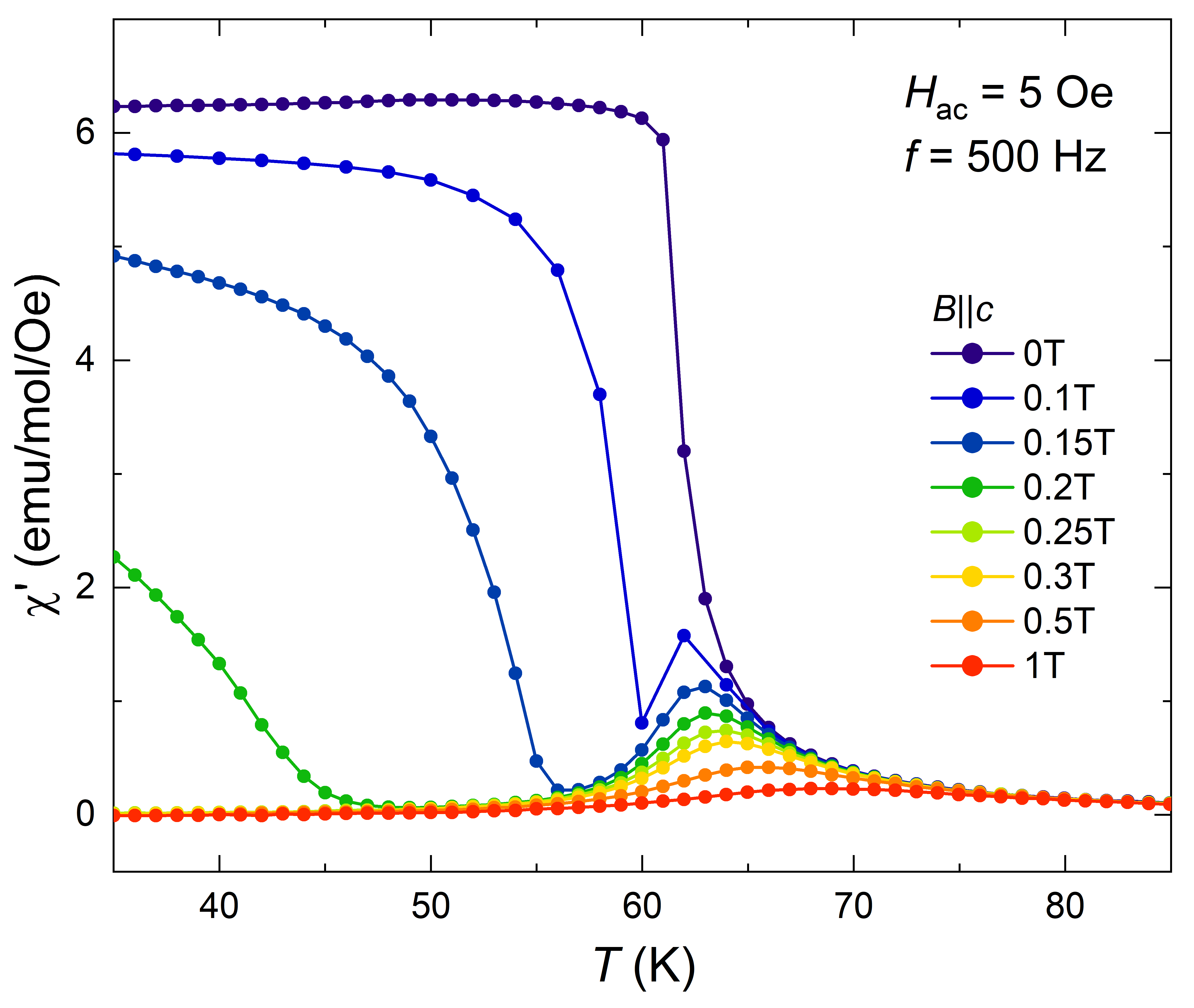}
    \caption{Temperature-dependence of the real part of the dynamic magnetic susceptibility at selected static magnetic fields applied $B\parallel c$. Measurements have been performed at an excitation frequency of $f = 500$~Hz.}
    \label{fig:supp_ACChi}
\end{figure}

The real part of the dynamic magnetic susceptibility measured with a small excitation field $H_\mathrm{ac} = 5$~Oe oscillating at $f = 500$~Hz is depicted in Fig.~\ref{fig:supp_ACChi} for different static magnetic fields applied along the $c$ axis. At $B = 0$~T, upon cooling from high temperatures, $\chi'$ is characterized by a steep increase at $T_\mathrm{C} = 61$~K marking the onset of ferromagnetic long-range order of the bulk phase. In contrast, $T_\mathrm{C}$ demonstrates itself as a maximum in $\chi'$ at finite magnetic fields, while a steep increase at lower temperatures signals the dynamics of the spin fluctuations. With increasing magnetic field, the crossover to the ferromagnetic regime '$T_\mathrm{C}$' slightly shifts to larger temperatures as expected for a ferromagnetic ground state.

\section{\textit{Ab-initio} Protocol}\label{sec:a2}
Strain structures are prepared and relaxed in \textit{ab-initio} VASP software package \cite{forvasp1,forvasp2}, version 6.5.1, which uses the projector augmented wave (PAW) basis set. A plane-wave energy cutoff of 600eV is used, and momentum space is sampled on a $12\times12\times12$ Gamma-centred grid.
For the exchange-correlation functional we use the general gradient approximation (GGA) as implemented by the revised Perdew-Burke-Ernzerhof GGA (PBEsol) method \cite{GGAPBEsol} for structural optimization and total energy calculations. 
On-site Coulomb interactions of the $3d$-orbitals of $\mathrm{Cr}$ are accounted for by use of the Dudarev simplified rotationally invariant DFT+U formalism \cite{DudarevDFTU} with an effective $U_{\text{eff}}=2\text{eV}$. 
Relaxations are carried out with a force stopping criterion of $0.5\cdot 10^{-3} \text{eV/\AA}$, and performing full relaxation, while keeping the conventional $c$-axis fixed to simulate strain. During the relaxation a, FM order is assumed, and the FFT grids are set constant across all runs.
	
The resulting structures are fed into ab-initio FPLO software package \cite{fplo1,fplo2}, version 22.00-62, for wannierization. We use the similar settings like above, but stay within the bare GGA, in order to estimate the kinetic terms without the correlations. Wannierization is carried out on $3d$-like Cr centred orbitals and $4p$-like I centred orbitals, on a large enough energy window $-7\,\mathrm{eV}$ to $3\,\mathrm{eV}$, allowing us to resolve separately direct and indirect hopping channels among the metal and ligand sites. 

\subsection{Strain relaxation results}
	\begin{figure}[ht]
 		\centering
       \begin{overpic}[width=0.5\columnwidth,percent,grid=false,tics=2]{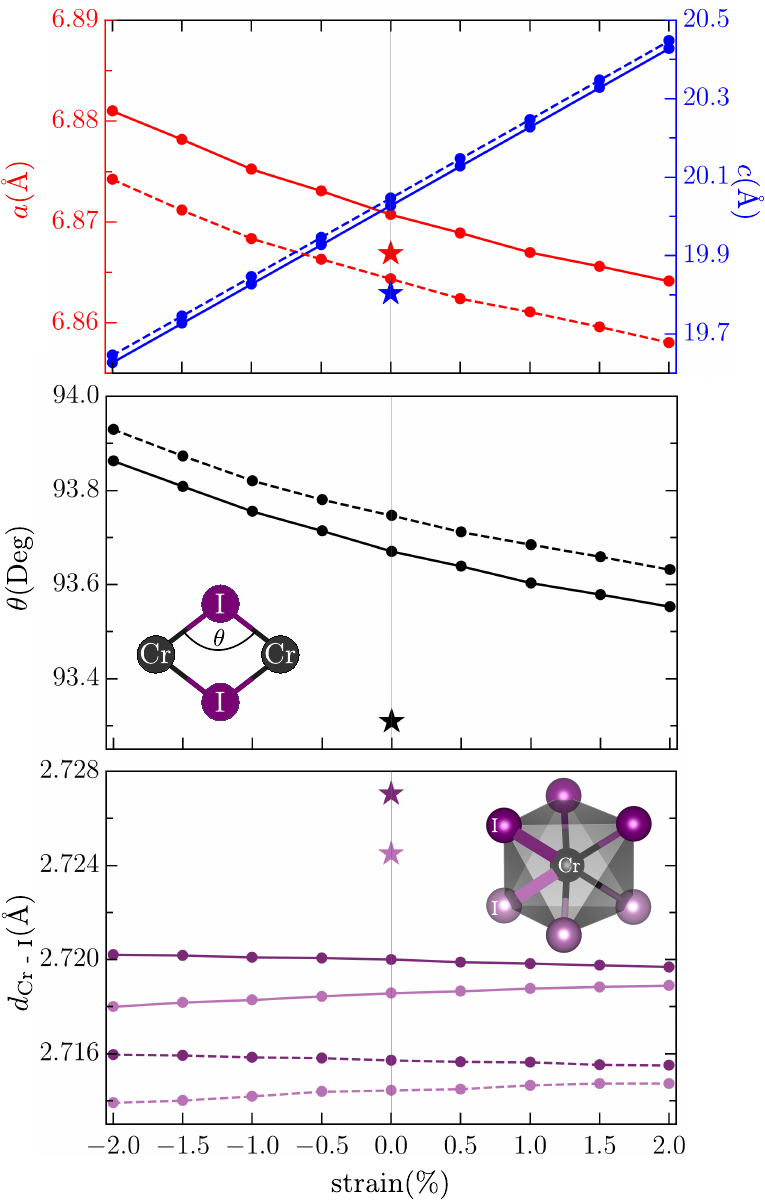}
        \put(9.5,96.5){(a)}
        \put(9.5,65){(b)}
        \put(9.5,33.6){(c)}
       \end{overpic}  
		\caption{Structual details from strain relaxations. The original seed structure, as reported in Ref~[\citenum{mcguire2015}] is shown as stars. Solid lines are from relaxations as presented in the computational details, and dashed lines are from the same settings but additionaly turning on SOC. (a) Lattice constants $a$ and $c$ as a function of strain.(b) The angle Cr-I-Cr as a function of strain. (c) The two independent distances Cr-I allowed by the $R\overline{3}$ spacegroup.
        }
		\label{fig:struct_relx_res}
	\end{figure}
	
The structural summary of the relaxations is seen in Fig.~\ref{fig:struct_relx_res}, with results from the above settings, as well as an additional relaxation where we also turn on SOC. We see that with and without SOC there is no appreciable change. During the relaxation we have kept the conventional $c$-axis fixed, seen in panel (a) as a perfectly straight line. The in-plane constant $a$ responds to this and relaxes, as do also the atom co-ordinates. Within $1\%$ strain we are still within more or less linear response as seen from the plots. Comparing to the reported bulk \cri\ structure \cite{mcguire2015}, plotted as stars in the figure, we see that the ambient \textit{ab initio} relaxation for zero strain matches relatively well.

An important microscopic structural scale is the super-exchange angle Cr-I-Cr. In panel (b) we see that the variation of the angle is not that significant over strain. We remain in an approximate $90^{\mathrm{o}}$ geometry. Within the R$\bar{3}$ spacegroup, the position of the I above and below Cr, within one octahedra cage (see inset of panel (c)) are not symmetry related and are free to be any value. Nevertheless, the chemistry works out such that distances are almost the same. This is seen in panel (c) where the two symmetry-inequivalent distances are almost identical, within $10^{-2}\text{\AA}$ difference. 

\subsection{Extracting magnetic exchanges}
Given the aforementioned structural results, we are within the framework used in Ref.~[\citenum{StavropoulosPRR2021}] to calculate the magnetic exchanges, by using the extracted metal to ligand hoppings $d-p$ as well as the on-site atomic energy difference between the metal $d$ and ligand $p$ manifolds, as found from wannierization. An effective I spin-orbit coupling $\lambda_p=0.5\text{eV}$ is used, and the Hubbard-Kanamori parameters are set to $U=5\text{eV}$, $J_H=1\text{eV}$, $U'=U-2J_H$. The resulting magnetic exchange couplings $\mathcal{J}$ of the nearest neighbor bonds are shown in Fig.~\ref{fig:MagnetoElastiCouplings}. 
While the bond magnetic exchanges results in reasonable values consistent with other predictions in the literature \cite{lee2026,cen2023DeterminingKitaevInteraction}, we note that SIA appears unreasonably large. Within the approximation of the effective down-folded perturbation procedure used, SIA is the most difficult to predict. This is because any and all metal-ligand and ligand-ligand processes can contribute non-trivially to setting SIA. Comparatively, the bond exchanges are set usually by a handful of spread-out ligands, in this case just two. Therefore any approximations or error in the estimation can accrue much more rapidly for the SIA, while bond exchanges are usually well predicted. 

To overcome this hurdle, we switch to a different way of estimating SIA by energy differences from Density Functional Theory (DFT). Here, we perform fully relativistic spin-polarized calculations within the FPLO framework~\cite{fplo1} at the level of GGA, where we compare two polarization directions: along the crystallographic $c$-axis and in the $ab$-plane. By computing the energy difference between the two polarization directions, $E_c-E_{ab}$, for each strained structure, we obtain the SIA as a function of strain as plotted in Fig. \ref{fig:MagnetoElastiCoupling_SIA}.

\section{Magnetoelastic Couplings}
The magnetoelastic couplings $\tilde{\mathcal{J}}$ are the derivatives of the magnetic exchange couplings $\mathcal{J}$ with respect to strain. In practice, these derivatives are determined via linear regression, where the slope is identified as the magnetoelastic coupling.
The linear regressions are performed using the \texttt{scipy.stats.linregression} package in python3 \cite{2020SciPy-NMeth}. {In Fig.~\ref{fig:MagnetoElastiCouplings} and \ref{fig:MagnetoElastiCoupling_SIA} we see along with the exchanges the fitted linear regression.
\begin{figure}[!h]
    \centering
    \includegraphics[width = 0.5\columnwidth]{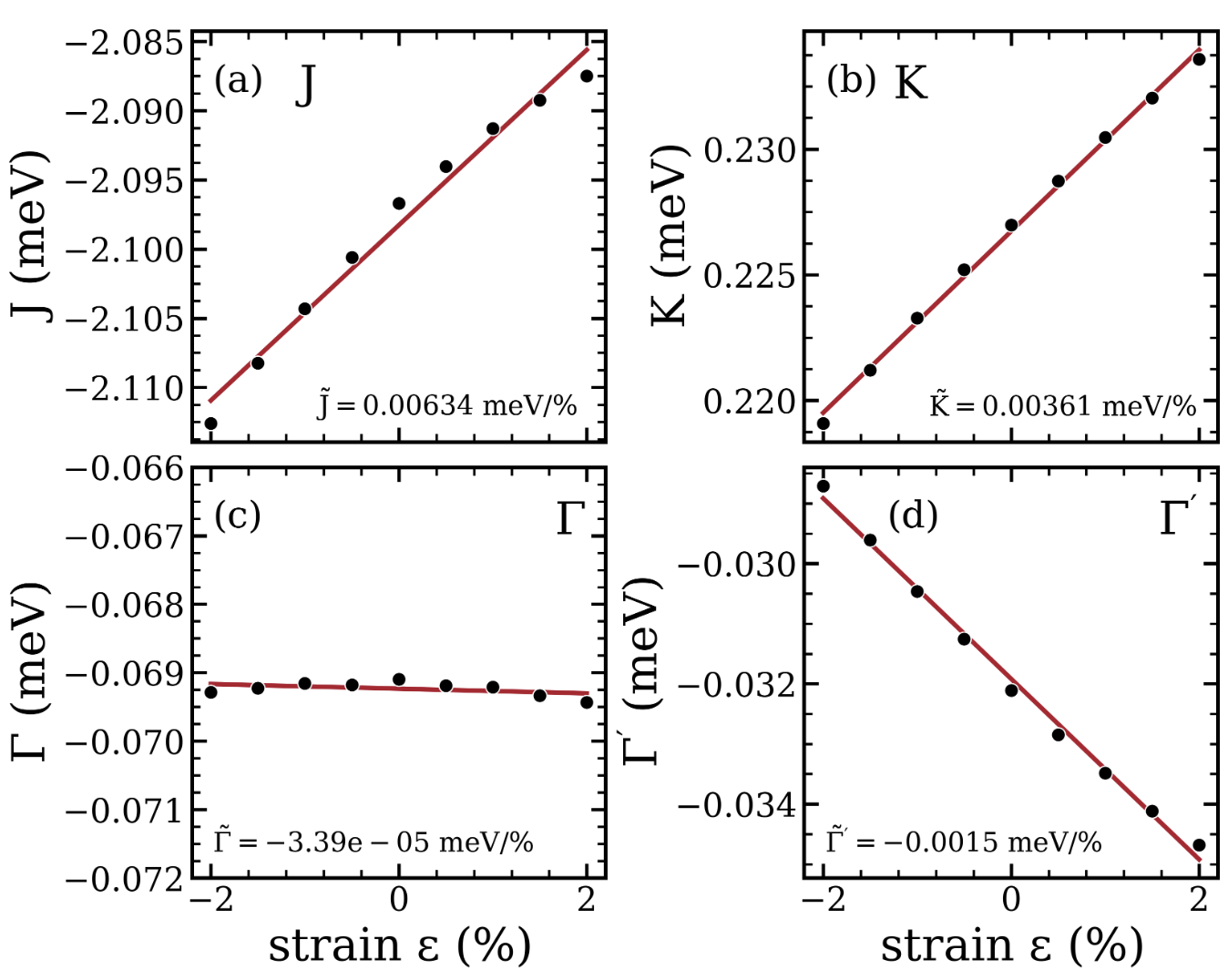}
        \caption{Magnetic couplings $\mathcal{J}$ obtained by the \textit{ab initio} protocol explained in the text for different strains. The slope of the fit of the couplings over strain denotes the magnetoelastic couplings $\tilde{\mathcal{J}}$.}
    \label{fig:MagnetoElastiCouplings}
\end{figure}
\begin{figure}[!h]
        \centering
        \includegraphics[width=0.5\columnwidth]{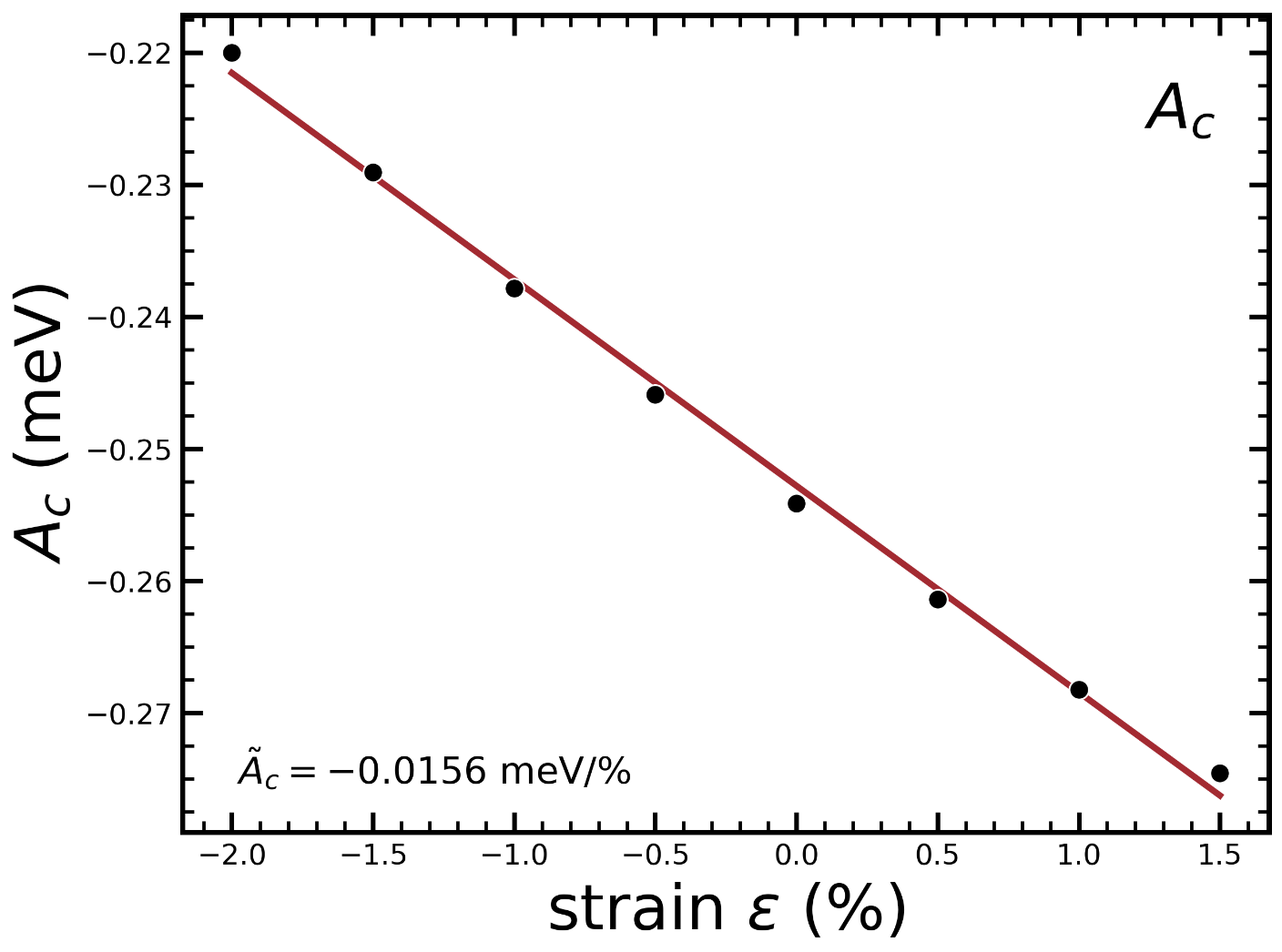}
        \caption{Single Ion Anisotropy $A_c$ obtained by DFT. See Fig.~\ref{fig:MagnetoElastiCouplings} for corresponding details.}
        \label{fig:MagnetoElastiCoupling_SIA}
\end{figure}
\section{Self-consistent Cluster Mean Field ED Calculation}
We employ a self-consistent cluster mean-field theory method (SCCMFT), in which the intralayer interactions ($H_0$) are treated on a fully-quantum level by exact diagonalization of eight-site honeycomb clusters with periodic boundary conditions. 

The interlayer interactions ($H_\perp$) are incorporated on mean-field level, $H_\perp\approx J_{\perp} \sum_{\langle ij\rangle_\perp} \mathbf S_i\cdot \langle\mathbf S_j\rangle$, where $\langle\mathbf S_j\rangle$ on neighboring layers are determined by the self-consistency loop outlined below. We chose a supercell of two 8-site honeycomb clusters, such that the A-type antiferromagnetism can be hosted on the $2\times8$ site supercell. The Hamiltonian is written as $H = H_0 + H_{\perp}$, where \(H_0\) collects all intralayer terms and \(H_{\perp}\) describes the interlayer interactions by an effective field. We diagonalize \(H_0+H_{\perp}\) for one surface layer, compute the spin expectation value to define the mean-field Hamiltonian acting on the other layer. Iterating this procedure between the two coupled layers yields a fixed point where the spins become stationary. This self-consistent solution incorporates quantum correlations exactly within each layer while treating the interlayer couplings at the static mean-field level.\\
For the two layers, labeled by \(A\) and \(B\), with the respective Hamiltonians \(H_A\) and \(H_B\), coupled via the interlayer exchange $J_{\perp}$, the self-consistent mean fields are obtained iteratively as follows:
\begin{enumerate}
    \item Solve layer \(A\) with the bare cluster Hamiltonian \(H_A\) and determine its ground state \(|\Psi_A\rangle\),
\begin{align}
    H_0 \, |\Psi_A\rangle = E_A |\Psi_A\rangle\,.
\end{align}
    \item Measure the spin expectation value on layer \(A\),
\begin{align}
    \braket{\mathbf{S}_j^{A}} \equiv \langle \Psi_A | \mathbf{S}_j^{A} | \Psi_A \rangle .
\end{align}

    \item Use these expectation values as input to construct the mean-field term acting on layer \(B\),
\begin{align}
    H_{\perp}^{A \rightarrow B}\!\left[\braket{\mathbf{S}_j^{A}}, J_\perp\right]
    \;=\;
    J_{\perp} \sum_{\langle ij\rangle_\perp} \, \braket{\mathbf{S}_j^{A}} \mathbf{S}_i,
\end{align}
    where \(J_\perp\) denotes the effective interlayer coupling entering the surface SCCMFT description.

    \item Solve layer \(B\) with the corresponding mean-field Hamiltonian and obtain the ground state \(|\Psi_B\rangle\),
\begin{align}
    \left(H_0 + H_{\perp}^{A \rightarrow B}\right)|\Psi_B\rangle = E_B |\Psi_B\rangle\,.
\end{align}
    \item Measure the local spin expectation values on layer \(B\),
\begin{align}
    \braket{\mathbf{S}_j^{B}} \equiv \langle \Psi_B | \mathbf{S}_j^{B} | \Psi_B \rangle .
\end{align}
    and use them to construct the mean-field term acting back on layer \(A\),
\begin{align}
    H_{\perp}^{B \rightarrow A}\!\left[\braket{\mathbf{S}_j^{B}}, J_\perp\right]
    \;=\;
    J_{\perp} \sum_{\langle ij\rangle_\perp} \, \braket{\mathbf{S}_j^{B}} \mathbf{S}_i,
\end{align}
    \item Solve layer \(A\) again with the updated Hamiltonian
\begin{align}
    \left(H_0 + H_{\perp}^{B \rightarrow A}\right)|\Psi_A\rangle = E_A |\Psi_A\rangle,
\end{align}

    \item Repeat steps \(2\)--\(6\) iteratively until the mean-field terms
\begin{align}
    H_{\perp}^{B \rightarrow A}, \qquad H_{\perp}^{A \rightarrow B},
\end{align}
    or equivalently the moments $\braket{\mathbf{S}_j^{A}}$ and $\braket{\mathbf{S}_j^{B}}$, no longer change within a chosen numerical tolerance $\varepsilon$.
\end{enumerate}
At convergence, the resulting pair of states \((|\psi_A\rangle,|\psi_B\rangle)\) define the SCCMFT solution for the two coupled layers. All presented calculations have been carried out with $\varepsilon = 10^{-13}$.
\section{Contributions to Length change}
\label{sec:contributionsLengthChange}
To analyze the magnetoelastic response in greater detail, we decompose the magnetostriction into its individual contributions. 
The in-plane results are labeled $ab$ while strictly being results obtained in the $b$ direction as discussed in the main text. However, we also computed the corresponding quantities for fields applied along $a$ and found qualitative agreement. Thus, we label the $b$ results as $ab$.

We assume a bulk fraction of $x = 0.9$, such that the total expectation value of an observable $\mathcal{O}$ is given by its bulk and surface contributions via
\begin{align}
\langle \mathcal{O} \rangle = x\, \langle \mathcal{O} \rangle_{\mathrm{bulk}} + (1-x)\, \langle \mathcal{O} \rangle_{\mathrm{surface}}\,.
\end{align}
As described in the main text, the magnetostriction coefficient is approximated as
\begin{align}
        \frac{1}{L_i}\left. \frac{\partial L_i}{\partial B_j}\right|_{p_i} &= 
        \frac{1}{V} \left. \frac{\partial M_j}{\partial p_i}\right|_{B_j}\\
        &\approx 
        \frac{\kappa_i \,L_i }{V}\,
         \sum_{\mathcal J\in\{J,\,K,\, \Gamma',\dots\}}
         \frac{\partial \mathcal J}{\partial L_i} 
         \left. \frac{\partial M}{\partial \mathcal J} \right|_{B_j} ,\label{eq:magnetostrictiontotal}
\end{align}
where for in-plane directions $i={ab}$ we assumed $\frac{\partial \mathcal J}{\partial L_c}=-C \frac{\partial \mathcal J}{\partial L_{ab}}$ with $C$ as a unitless positive constant, while ${L_c}\frac{\partial \mathcal J}{\partial L_c}=\tilde{\mathcal{J}}$ were obtained from \textit{ab initio}. The susceptibilities $\frac{\partial M}{\partial \mathcal J} $ were obtained in the SCCMFT calculations using a Maxwell relation
\begin{equation}
    \frac{\partial M}{\partial \mathcal{J}}
    =
    - \frac{\partial \langle H_{\mathcal{J}} \rangle}{\partial B},
    \qquad
    H_{\mathcal{J}} \equiv \frac{\partial H}{\partial \mathcal{J}},
    \label{eq:maxwell_M_J}
\end{equation}
where $H_{\mathcal{J}}$ denotes the part of the Hamiltonian related to a particular coupling $\mathcal{J}$.
The form of Eq.~\ref{eq:magnetostrictiontotal} allows to dissect the magnetostriction into summand contributions from magnetoelastic couplings of different types 
$\mathcal J \in \{J,K,\Gamma,\Gamma',A_c,\dots \}$. 
\begin{figure*}[ht]
    \centering
    \includegraphics[width=\linewidth]{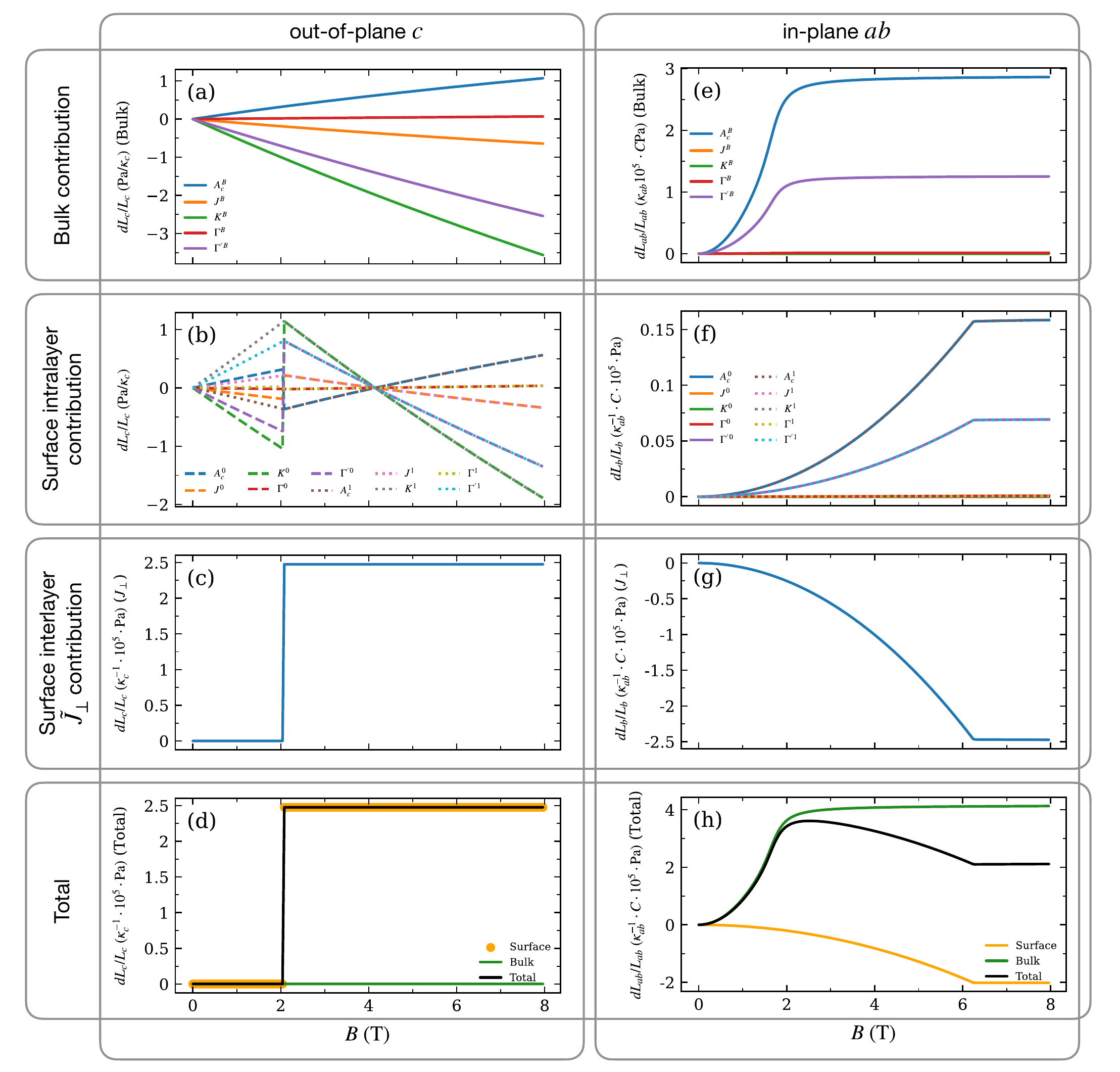}
    \caption{Contributions to the length change $dL_i/L_i$ in (in-plane) $b$ direction (a-d) and (out-of-plane) $c$ direction (e-h) for bulk (a,e), intralayer contributions of the two surface layers (b,f), as well as the interlayer $J_{\perp}$ contribution (c,g) and the combined results replotted from the main text (d,h). The superscripts denote the bulk $\mathcal{J}^B$ and surface contributions for the layers labeled layer $0$ $\mathcal{J}^0$ and layer 1 $\mathcal{J}^1$, plotted solid and dashed respectively.}
    \label{fig:contributions_magnetostriction}
\end{figure*}
This dissection is shown in detail in Fig.~\ref{fig:contributions_magnetostriction}, where the left (right) panel considers out-of-plane (in-plane) length change and field, and the rows from top to bottom correspond to the contributions by bulk, the intralayer surface contributions for each of the two layers plotted solid and dashed, respectively, the surface interlayer coupling $J_{\perp}$, and the total summed result (equal to that shown in the main text), respectively. 

We first discuss the out-of-plane magnetostriction contributions displayed in the left column of Fig.~\ref{fig:contributions_magnetostriction}. 
For the length change in $c$ direction (easy-axis direction), the situation is qualitatively similar to the magnetization curve: the critical field $B^{\star}$ causes a jump in the magnetization and the sudden change from AFM to FM surface layers couples most strongly to $J_\perp$. This behavior is consistent with the fact that each layer remains nearly fully polarized due to the strong single-ion anisotropy $A_c$. 

Consequently, all contributions except for $J_{\perp}$ are approximately constant. This can be seen in more detail in Fig.~\ref{fig:contributions_magnetostriction}(c) displaying the $J_{\perp}$ contribution only and Fig.~\ref{fig:contributions_magnetostriction}(d) depicting the surface and bulk contributions as plotted in the main text. In order to highlight the nature of the only-surface contributions, the surface has been highlighted in orange to be set visually apart from the total result.

We now focus on the contribution of the bulk alone which is plotted in Fig.~\ref{fig:contributions_magnetostriction}(a). Due to the bulk being nearly polarized, the contributions remain constant and of little magnitude, the magnitude of the signals in panel (b) is on the order of $10^{-5}$ smaller than the surface (please note the units).

The same logic applies to the contribution of the surface intralayer couplings plotted in Fig.~\ref{fig:contributions_magnetostriction}(b). The surface layers are also nearly polarized for the whole field range and of $10^{-5}$ magnitude smaller than the other contributions. Here, the two contributions of the simulated layers are plotted as solid and dashed lines. 

The overall structure of the intralayer surface contribution curves, however, looks completely different to the bulk contribution: one can see a flip of sign at the critical field $B^{*}$ at $B \approx 2.1\,\mathrm{T}$  in the contributions of both layers as consequence of the flip of one layer's magnetic moment to also align with the magnetic field direction. The mirrored sign structure for $B < 2.1\,\mathrm{T}$ can be explained as follows: For zero applied field, both layers are already almost, but not entirely, polarized. The magnetization in one layer is oriented parallel and the other layer antiparallel to the $c$-axis due to the AFM $J_{\perp}$ incorporated via SCCMFT. When a field is applied, the magnetization is slightly increased (decreased) in the layer aligned (antialigned) with the field direction.

Having dissected the out-of-plane contribution, we now analyze the in-plane $ab$ contributions displayed in the right column of Fig.~\ref{fig:contributions_magnetostriction}. As stated in the main text and displayed in Fig.~\ref{fig:contributions_magnetostriction}(h), the increase and decrease in $dL_{ab}/L_{ab}$ is caused by interplay between the bulk contribution displayed in (e) and the interlayer coupling $J_{\perp}$ shown in (g), while the surface intralayer contributions displayed in (f) are comparatively small.

The $ab$ bulk contribution is plotted in Fig.~\ref{fig:contributions_magnetostriction}(e). The dominant magnetoelastic couplings are the SIA $\tilde A_c$ as well as the anisotropic $\tilde \Gamma'$. The large magnetoelastic contribution of $\tilde \Gamma'$ is especially surprising as the zero-strain magnetic $\Gamma'$-exchange is negligible. 
Same applies for the surface layer contributions as plotted in Fig.~\ref{fig:contributions_magnetostriction}(f) where both simulated layers give the exact same contribution. 
The contribution related to $\tilde J_{\perp}$ as plotted in Fig.~\ref{fig:contributions_magnetostriction}(g) is the driving force of the decrease observed in the in-plane length change. This interplay explains the structure of the final result displayed in Fig.~\ref{fig:contributions_magnetostriction}(h).

To elucidate the pronounced contributions related to $\tilde A_c$ and $\tilde \Gamma'$ for a magnetic field applied in the $ab$ plane, we consider a minimal classical picture. 
At zero field strengths, the magnetic moments are parallel to the $c$ direction due to the strong single-ion anisotropy.  The increasing magnetic field along $ab$ tilts these toward the $ab$ direction. 
As shown in Fig.~\ref{fig:supp_HJ}  for the classical polarized state, the individual interaction energies $\braket{H_\mathcal{J}}$ couple strongly differently to different couplings $\mathcal{J}$, when rotating the polarized state from out-of-plane $c$ to in-plane $b$ direction: $\braket{H_{\Gamma^{'}}}$, $\braket{H_{\Gamma}}$ and $\braket{H_{A_c}}$ all couple to this rotation significantly, i.e.\ they are strong easy-plane/easy-axis couplings. To the length change in Eq.~\ref{eq:magnetostrictiontotal}, they contribute only as products with the magnetoelastic couplings $\tilde{\Gamma}'=-0.15\,\mathrm{meV}$, $\tilde{\Gamma}=-0.00\,\mathrm{meV}$, $\tilde{A_c}=-1.56\,\mathrm{meV}$, which is the reason why the $\tilde{A_c}$ and $\tilde{\Gamma}'$ contributions dominate in the end. 

\begin{figure}[h]
    \centering
    \includegraphics[width = 0.5\textwidth]{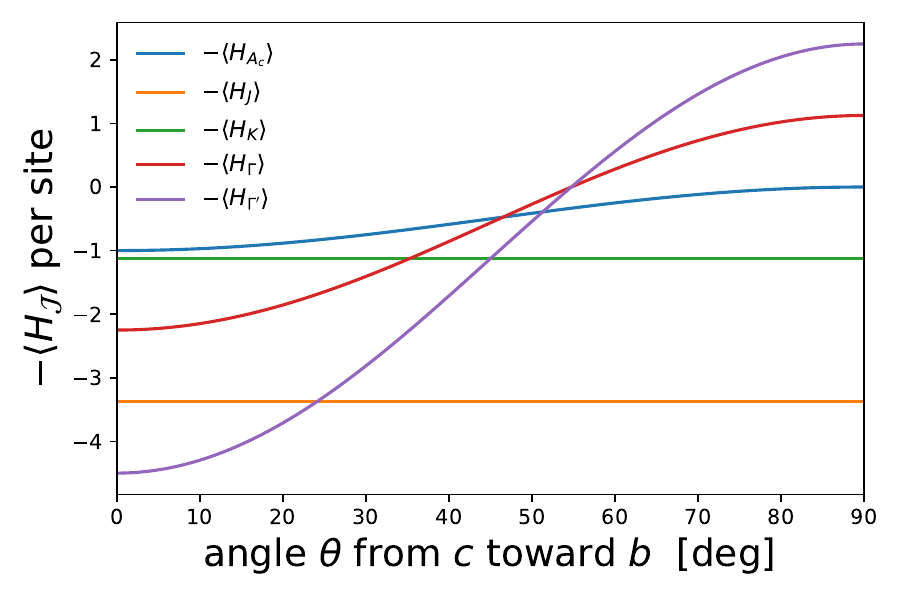}
    \caption{Contributions $-\langle H_J \rangle$ from each coupling for a spin rotation from $c$- to $b$-axis on the classical level.}
    \label{fig:supp_HJ}
\end{figure}

\bibliography{Bib_CrI3_MS}

\end{document}